\documentclass[pra,showpacs,twocolumn,superscriptaddress,preprintnumbers,eqsecnum,notitlepage,longbibliography]{revtex4-2}
\pdfoutput=1
\usepackage[utf8]{inputenc}
\usepackage[bookmarks=false,colorlinks=true,urlcolor=blue,citecolor=blue,linkcolor=blue]{hyperref}
\usepackage{natbib}

\usepackage{amsmath,amssymb,mathtools}
\usepackage{graphicx}
\usepackage{bm,color}
\usepackage[]{cleveref}
\usepackage{multirow}
\crefname{figure}{Fig.}{Figs.}

\newcommand{\be}{\begin{equation}}
\newcommand{\ee}{\end{equation}}
\newcommand{\nn}{\nonumber}

\newcommand{\sket}[1]{| #1 \rangle}

\newcommand{\ssandwich}[3]{\langle #1 | #3 | #2 \rangle}

\newcommand{\edit}[1]{#1}

\begin{document}

\preprint{FERMILAB-PUB-21-387-QIS}
\title{Benchmarking variational quantum eigensolvers for the square-octagon-lattice Kitaev model}

\author{Andy C.~Y.~Li}
\affiliation{Fermi National Accelerator Laboratory, Batavia, IL, 60510, USA}
\author{M. Sohaib Alam}
\affiliation{Rigetti Computing, Berkeley, CA, 94701, USA}
\author{Thomas Iadecola}
\affiliation{Department of Physics and Astronomy, Iowa State University, Ames, Iowa 50011, USA}
\affiliation{Ames National Laboratory, Ames, Iowa 50011, USA}
\author{Ammar Jahin}
\affiliation{Department of Physics, University of Florida, 2001 Museum Rd, Gainesville, FL 32611, USA}
\author{Joshua Job}
\affiliation{Lockheed Martin Advanced Technology Center, Sunnyvale, CA 94089}
\author{Doga Murat Kurkcuoglu}
\affiliation{Fermi National Accelerator Laboratory, Batavia, IL, 60510, USA}
\author{Richard Li}
\affiliation{Department of Physics, Yale University, New Haven, Connecticut 06520, USA}
\author{Peter P. Orth}
\affiliation{Department of Physics and Astronomy, Iowa State University, Ames, Iowa 50011, USA}
\affiliation{Ames National Laboratory, Ames, Iowa 50011, USA}
\author{A. Barış Özgüler}
\affiliation{Fermi National Accelerator Laboratory, Batavia, IL, 60510, USA}
\author{Gabriel N. Perdue}
\affiliation{Fermi National Accelerator Laboratory, Batavia, IL, 60510, USA}
\author{Norm M. Tubman}
\affiliation{Quantum Artificial Intelligence Lab. (QuAIL), Exploration Technology Directorate, NASA Ames Research Center, Moffett Field, CA 94035, USA}
\date{\today}

\begin{abstract} 
Quantum spin systems may offer the first opportunities for beyond-classical quantum computations of scientific interest.
While general quantum simulation algorithms likely require error-corrected qubits, there may be applications of scientific interest prior to the practical implementation of quantum error correction.
The variational quantum eigensolver (VQE) is a promising approach to finding energy eigenvalues on noisy quantum computers.
Lattice models are of broad interest for use on near-term quantum hardware due to the sparsity of the number of Hamiltonian terms and the possibility of matching the lattice geometry to the hardware geometry.
Here, we consider the Kitaev spin model on a hardware-native square-octagon qubit connectivity map, and examine the possibility of efficiently probing its rich phase diagram with VQE approaches.
By benchmarking different choices of variational ansatz states and classical optimizers, we illustrate the advantage of a mixed optimization approach using the Hamiltonian variational ansatz (HVA) and the potential of probing the system's phase diagram using VQE.
We further demonstrate the implementation of HVA circuits on Rigetti's Aspen-9 chip with error mitigation.
\end{abstract}

    \maketitle

\section{Introduction}

In the context of quantum computation there is reason to believe the quantum simulation of spin systems may offer early results in the search for beyond classical computations of real scientific interest \cite{Childs9456}.
Although beyond-classical calculations that offer truly new insights into scientific problems will likely require quantum error correction (QEC) \cite{Shor1995,Roffe2019}, it may be possible to find approachable questions of scientific interest even before the advent of full QEC if we carefully pair the computing hardware and problem. Lattice models are a natural class of systems to consider in this regard, and they also tend to have sparse Hamiltonian representations that require fewer quantum resources than other Hamiltonian models~\cite{hubbard2020}.
In this paper, we consider a spin model \cite{Kitaev2006,Yang07} that maps naturally onto the square-octagon qubit connectivity of a superconducting quantum processor and perform an open study of the most efficient ways of finding the ground state of this system using variational quantum algorithms based on parameterized quantum circuits. 

\begin{figure*}
	\centering
	\includegraphics[width=.95\textwidth]{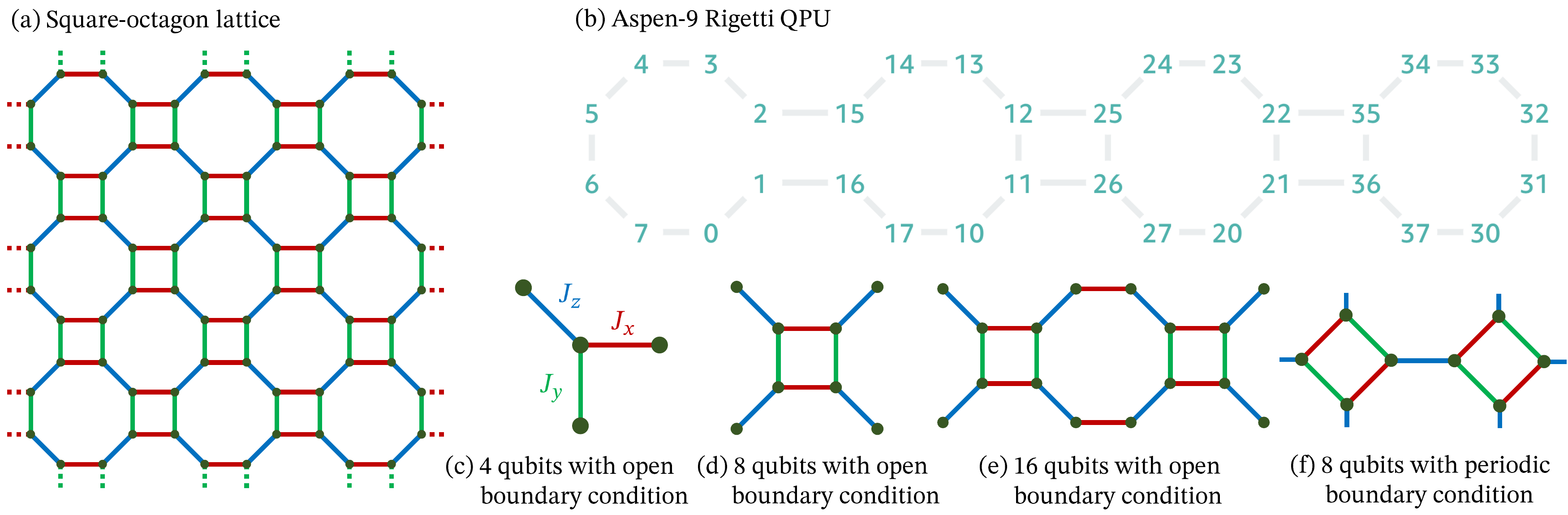}
	\caption{\textbf{Lattice geometry}. A square-octagon lattice with the Kitaev couplings is shown in (a). The Aspen-9 QPU is arranged with a square-octagon connectivity shown in (b). Note that qubits 1 and 2, and qubits 15 and 16 are not connected due to a hardware issue. The Aspen-9 geometry allows us to pick an appropriate sublattice to simulate a square-octagon-lattice Kitaev model with open boundary conditions, for example, the 4-qubit setup in (c), the eight-qubit lattice in (d) and the 16-qubit lattice in (e).
	We also consider an eight-qubit setup with periodic boundary conditions, shown in (f), to demonstrate VQE calculations of expectation values.
	The $J_x$ (red), $J_y$ (green) and $J_z$ (blue) couplings are assigned to each linkage.
	}
	\label{fig:lattice_geometry}
\end{figure*}

The near-term, so-called ``Noisy Intermediate Scale Quantum'' (NISQ) \cite{Preskill2018quantumcomputingin, cerezo2020variational, bharti2021noisy} computing era is defined by the conditions imposed by noisy quantum hardware. 
Decoherence errors mandate the use of ``shallow'' quantum circuit programs, i.e., quantum circuits with a small number of consecutive operations on the qubit register array.
Quantum noise also currently limits the effective ``width'' of a circuit, which is governed by the total number of qubits deployed. A wide circuit consists of more quantum gates than a narrow circuit, and thus the gate error rates effectively limit how wide a circuit could be before the result becomes no longer useful with the decreasing circuit fidelity.

The variational quantum eigensolver (VQE) \cite{peruzzoVariationalEigenvalueSolver2014,O'Malley2016,mcclean2016,Kandala2017,grimsley2019,huggins2020,Zhang-PRB-2021,tang2021,ctrlvqe} is an algorithm that seeks to implement low-depth variational ansatz circuits to find eigenstates of complex many-body Hamiltonians. Different approaches to VQE entail different choices of ansatz circuits, which take into account different figures of merit. These can include reducing the circuit depth, increasing the accuracy or wavefunction fidelity, or minimizing the number of variational parameters. To match these varied goals, there are many different approaches which include hardware efficient ansaetze (HEA) \cite{Kandala2017}, adaptive ansaetze~\cite{grimsley2019,tang2021, Zhang-PRB-2021,Gomes-AdvQuTech-2021,vanDyke-arxiv-2022,Mukherjee-arXiv-2022,Anastasiou-arXiv-2022}, and ansaetze inspired by classical simulations~\cite{Kandala2017,tang2021,clustervqe}. One particular ansatz of interest is the Hamiltonian variational ansatz (HVA) \cite{Wecker2015}.  This ansatz draws from ideas expressed by the quantum approximate optimization algorithm and adiabatic quantum computation~\cite{farhi2000,farhi2014}.
Recent work \cite{Wiersema2020} suggests the HVA may be less prone to problems with ``barren plateaus'' \cite{McClean2018} and therefore easier to optimize than the HEA (see, however, Ref.~\cite{larocca2021diagnosing}).

Kitaev spin models~\cite{Kitaev2006,Hermanns18,Takagi19}, which are a family of frustrated quantum spin models with bond-dependent interactions, provide an intriguing testbed for NISQ quantum simulation. Kitaev models can be defined on arbitrary trivalent graphs and are appealing due to their exact solvability via a mapping to free Majorana fermions~\cite{Kitaev2006}. Despite their simplicity, they yield a rich variety of phases, including gapped $\mathbb Z_2$ and gapless U(1) spin liquids~\cite{Hickey19}. In a magnetic field, Kitaev models support non-Abelian Ising anyons~\cite{Kitaev2006}, which (in addition to the aforementioned $\mathbb Z_2$ spin-liquid phase) 
makes them a promising platform for topological quantum computation~\cite{Xu11,Lee17,Freedman2002}. Kitaev-like models are predicted to arise in spin-orbit-coupled Mott insulators~\cite{Jackeli09}, and a plethora of materials candidates exist~\cite{Takagi19}. Putative signatures of spin-liquid physics have been observed in neutron scattering~\cite{Banerjee16,Banerjee_2018} and thermal transport measurements~\cite{Kasahara18,Yokoi20}.

In realistic systems, the desired ``Kitaev interactions" compete with more mundane (e.g. Heisenberg) interactions and external magnetic fields, all of which spoil the Kitaev model's exact solvability and often favor magnetically ordered ground states. One therefore must resort to numerical methods~\cite{Chaloupka10,Jiang2011,Chaloupka2013,Osorio14,Rau14,Shinjo15,Winter16,Yadav16,Gohlke17,Gotfryd17,Zhu18,Patel19,Gordon19,Czarnik19,Lee20}, such as exact diagonalization (ED), tensor-network techniques like the density-matrix renormalization group~\cite{White92,Schollwock11,fishman2020}, and Monte Carlo~\cite{mishchenko2021quantum, Kurita_2015}, to study the ground-state phase diagram. Thus, Kitaev models provide a useful benchmark for near-term quantum algorithms for the study of interacting quantum systems.

In this paper, we focus on the Kitaev model on the square-octagon lattice~\cite{Yang07,Kells11}, which maps natively with minimal compilation overhead onto Rigetti's Aspen architecture featuring 32 transmons with a square-octagon topology on the Aspen-9 chip used in this paper~\cite{PhysRevA.101.012302,Abrams2020,Reagor2018}.
This native mapping makes the quantum simulation of the square-octagon Kitaev model a potential NISQ application of the Aspen-9 quantum processing unit (QPU) to carry out quantum computing calculations without full QEC.
We will benchmark the variational ansaetze and optimization algorithms to determine the ground state of the Kitaev model with and without a magnetic field. Supported by the results of classical simulations, we will illustrate a mixed optimization approach using both a local optimizer and a non-local optimizer started with multiple initial values.
This opens up the possibility of using VQE calculations to probe the phase diagram of the Kitaev model in the presence of a magnetic field or other perturbations away from the solvable limit.
Together with an experimental test run of classically optimized VQE circuits on the QPU, this paper provides insights into the appropriate VQE approach for further VQE experiments with a system size beyond the capability of classical computation.

This paper is organized as follows. In \cref{subsec:review_Kitaev_model}, we review the Kitaev model on the square-octagon lattice and its phase diagram in the presence of a magnetic field. This is followed by a review of VQE approaches in \cref{subsec:review_VQE}. We discuss our methodology for investigating the circuit ansaetze (\cref{subsec:circuit_ansatz}) and optimizers (\cref{subsec:optimizers}), and for implementing the classically optimized circuits on Aspen-9 with noise mitigation (\cref{subsecn:expt}). The results of the paper are then discussed in \cref{sec:results}.

\section{Background}
\label{sec:background}
\subsection{Square–octagon-lattice Kitaev model}
\label{subsec:review_Kitaev_model}

\begin{table}
	\caption{\textbf{Definition of studied model parameter values}. This table contains the model parameter values that are used in our benchmark simulations below. They correspond to positions in the phase diagram in the toric code phase without a magnetic field (TC$_z$) and with a magnetic field (TC$_z+h$), on the gapless line (GL) and above the gapless line in a magnetic field (GL+$h$). 
		The ground-state energy $E_g$ is determined by ED for the four-qubit setup ($N=4$), the eight-qubit lattice ($N=8$) and the 16-qubit lattice ($N=16$) with open boundary conditions shown in \cref{fig:lattice_geometry}(b), (c) and (d).}
	\medskip
	\centering
	\begin{tabular}{|c||c|c|c|c|c|c|| c | c | c|}
		\hline
		\multirow{2}{*}{Label} & \multicolumn{6}{|c||}{Model parameters } & \multicolumn{3}{|c|}{Ground-state energy $E_g$ }
		\\
		\cline{2-10} 
		& $J_x$ & $J_y$ & $J_z$ & $h_x$ & $h_y$ & $h_z$ & $N=4$ & $N=8$ & $N=16$ \\
		\hline
		\hline
		TC$_z$ & 0.1 & 0.1 & 1 & 0 & 0 & 0 & -1.0100 & -4.0100  &  -8.0250\\
		\hline
		TC$_z$+$h$ & 0.1 & 0.1 & 1 & $\frac{0.05}{\sqrt{3}}$ & $\frac{0.05}{\sqrt{3}}$ & $\frac{0.05}{\sqrt{3}}$ & -1.1723 & -4.2476 & -8.5002\\
		\hline
		GL & $\frac{1}{\sqrt{2}}$ & $\frac{1}{\sqrt{2}}$ & 1 & 0 & 0 & 0 & -1.4142 & -4.4721 & -9.3002\\
		\hline
		GL+$h$ & $\frac{1}{\sqrt{2}}$ & $\frac{1}{\sqrt{2}}$ & 1 & $\frac{0.05}{\sqrt{3}}$ & $\frac{0.05}{\sqrt{3}}$ & $\frac{0.05}{\sqrt{3}}$ & -1.5831 & -4.7011 & -9.7008 \\
		\hline
	\end{tabular}
	\label{tab:model_params}
\end{table}

\begin{figure}
\centering
    \includegraphics[width=\linewidth]{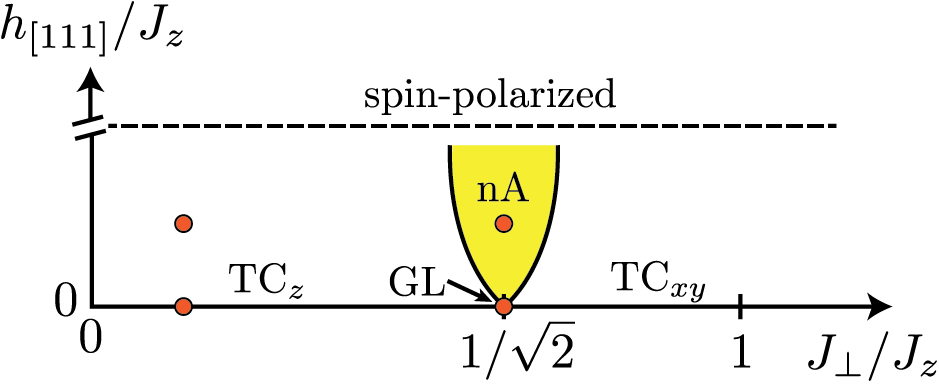}
    \caption{\textbf{Phase diagram}. Schematic phase diagram of the Kitaev model~\eqref{eq:H} on the square-octagon lattice as a function of spin exchange anisotropy $J_{\perp}/J_z$ with $J_\perp \equiv J_x = J_y$ and magnetic field in $[111]$ direction $h_{[111]}$. It includes gapped toric code phases (TC$_z$, TC$_{xy}$) that are stable with respect to small fields, the gapless line (GL) at $J_\perp/J_z = 1/\sqrt{2}$ and a phase with non-Abelian (nA) Majorana excitations that emerges in field above the gapless line. At large magnetic fields the system enters a spin-polarized paramagnetic phase. The red circles denote the different, representative model parameter points that are studied in our benchmark simulations.} 
    \label{fig:phase-diagram}
\end{figure}

The ferromagnetic Kitaev model on the square-octagon lattice has the Hamiltonian,
\begin{align}
	\label{eq:HK}
	\! \! \! \! H_{\rm K} = -J_x\sum_{\text{$x$-bonds}}X_{i}X_{j}-J_y\sum_{\text{$y$-bonds}}Y_{i}Y_{j}-J_z\sum_{\text{$z$-bonds}}Z_{i}Z_{j},
\end{align}
where $J_{x,y,z} > 0$ and we partition the nearest-neighbor bonds of the lattice into $x,y,$ and $z$ sets depending on their orientation (see Fig.~\ref{fig:lattice_geometry}(a)), and  $\{X_i,Y_i,Z_i\}$ are the Pauli matrices for site $i$. Like the original honeycomb-lattice version, the Kitaev model $H_{\rm K}$ on the square-octagon lattice is exactly solvable by mapping to free fermions~\cite{Yang07}. To move beyond the exactly solvable limit and explore the rich phase diagram emerging in finite magnetic field (see Fig.~\ref{fig:phase-diagram}), we consider the model
\begin{align}
	\label{eq:H}
	H = H_{\rm K} + \sum_i (h_x X_i+h_y Y_i +h_z Z_i),
\end{align}
where $\vec h=(h_x,h_y,h_z)$ is the magnetic field.
This Hamiltonian and variants thereof have been studied in Refs.~\cite{Yang07,Kells11,Berke20,Hickey21,Yamada21,Oleksandr_2021,Jahin_2022} using a variety of methods, and the general features of its phase diagram are well understood.

At zero field, the model features two gapped $\mathbb Z_2$ spin-liquid phases, which we dub $\mathrm{TC}_{z}$ and $\mathrm{TC}_{xy}$. The $\mathrm{TC}_{z}$ phase can be understood perturbatively in the limit $J_z^2\gg J_x^2+J_y^2$, where an effective Hamiltonian equivalent to the well-known toric code~\cite{Kitaev03} emerges~\cite{Yang07,Kells11}. The $\mathrm{TC}_{xy}$ phase can be understood in the opposite limit $J_x^2+J_y^2\gg J_z^2$, where perturbation theory yields~\cite{Kells11} a Hamiltonian equivalent to the so-called Wen-plaquette model~\cite{Wen03}, which is in turn equivalent to the toric code~\cite{Nussinov09,Brown11}. In both $\mathrm{TC}$ phases, the model has $\mathbb Z_2$ topological order and supports Abelian anyonic excitations with nontrivial mutual statistics; nevertheless, the phases are distinct~\cite{Kells11,Yamada21}. A critical line separating these two gapped phases appears at $J_z^2=J_x^2+J_y^2$.

At small but finite field, the $\mathrm{TC}$ phases persist~\cite{Hickey21}, as is expected due to the gapped nature of these phases. Provided that none of $h_{x,y,z}=0$, the gapless line at $J_z^2=J_x^2+J_y^2$ gives way to a gapped phase with non-Abelian Majorana excitations~\cite{Yang07}. At large field, the system enters a trivial spin-polarized paramagnetic phase. The phase diagram of the square-octagon Kitaev model in a field has been studied in Refs.~\cite{Yang07,Berke20,Hickey21}, but a detailed understanding of the location of all transitions is lacking. A schematic of the phase diagram for $J_x=J_y=J_\perp$ in a $[111]$-oriented field $\vec h=h_{[111]}(1,1,1)/\sqrt{3}$ is shown in Fig.~\ref{fig:phase-diagram}.

In the following, we focus on a number of representative points in the phase diagram, which are highlighted in Fig.~\ref{fig:phase-diagram} and defined in Table~\ref{tab:model_params}. The table also includes the exact ground-state energies for these parameters that are obtained using ED, which we will use to benchmark our VQE results. We consider lattices with open boundary conditions shown in \cref{fig:lattice_geometry}(b,c,d) for the benchmarks.

\subsection{VQE}
\label{subsec:review_VQE}
VQE algorithms prepare the eigenstates of the system Hamiltonian $H$ by optimizing the cost function associated with a trial state $\sket{\psi(\vec{\theta})}$, which is prepared by a parametrized circuit ansatz $U(\vec{\theta})$  such that $\sket{\psi(\vec{\theta})} = U(\vec{\theta}) \sket{0}$, where $\sket{0}$ is a chosen reference state \cite{O'Malley2016,Kandala2017}. To prepare the ground state, the VQE algorithm minimizes, using a classical optimizer, the energy of the trial state, i.e.,
\be
\label{eq:cost}
E(\vec{\theta}) = \ssandwich{\psi(\vec{\theta})}{\psi(\vec{\theta})}{H}
.
\ee
The cost function is measured on the QPU, and the result is fed into the classical optimizer, making the VQE a hybrid quantum-classical approach.
Several different approaches are possible for calculating excited states~\cite{peruzzoVariationalEigenvalueSolver2014,mcclean2016,varianceVQE,Zhang-PRB-2021}.  One approach is to minimize a modified cost function including the overlaps with the lower-energy eigenstates determined in the previous iterations \cite{Nakanishi2018,Higgott2019}.

Similar to other variational algorithms, the output of VQE only approximates the eigenstates of $H$, and the quality of the solution depends on the choice of ansatz.
The true ground state cannot be expressed by the trial state $\sket{\psi(\vec{\theta})}$ if an inappropriate ansatz is chosen, resulting in an approximated energy much higher than the true ground-state energy.
The success of VQE algorithms hence strongly depends on the circuit ansatz.
Optimizing a generic HEA is a challenging task for several reasons, including barren plateaus \cite{McClean2018} and sensitivity to the optimizer meta-parameters \cite{gu2021}.
%the vanishing gradients known as the Barren plateaus
% \cite{McClean2018,gu2021}
The challenge of optimization grows even bigger with the rapidly increasing number of variational parameters with the system size. The HVA, built using rotations generated by Hamiltonian terms, has been suggested to be a more efficient option than HEA in certain cases~\cite{Wiersema2020}, as has been observed in several numerical simulations \cite{Wen2019,Cade2020,gu2021}.
Later in this paper, we will compare the effectiveness of these two ansaetze in preparing the ground state of the square-octagon Kitaev model with 8 and 16 qubits.

The choice of classical optimization algorithm plays an important role in VQE implementations as well. Picking an inappropriate optimizer results in a slow convergence rate (and thus a longer QPU runtime), if it ever converges to a sufficiently good minimum at all \cite{Lavrijsen2020,Anand2021,harrow2021,gu2021}.
We will also benchmark the efficiency of a few optimizers using the Kitaev model to provide some insights into the VQE optimization strategy. 

\section{Methodology}
\subsection{Overview of approach}
In the following sections we review our theoretical and experimental approaches to studying VQE simulations of the Kitaev model.  Our paper first aims, via classical simulations, to understand the modeling requirements for obtaining different phases of the Kitaev model with various ansaetze and optimizers, and then tests these approaches on present-day quantum hardware.
In this paper, we classically simulate the quantum programs using PyQuil \cite{smith2016practical}, Cirq \cite{cirq_developers_2021} and Qiskit \cite{Qiskit2021}, and we execute classically optimized quantum circuits on Rigetti's Aspen-9 QPU.

\subsection{VQE Ansatz}
\label{subsec:circuit_ansatz}

\edit{
The QPUs are initialized in a chosen reference state $\sket{0}$ with all physical qubits being at zero in the computational basis. The ansatz trial state $\sket{\psi(\vec{\theta})} = U(\vec{\theta}) \sket{0}$ is prepared by two classes (HVA and HEA) of parametrized circuit ansatzes $U(\vec{\theta})$ to be discussed in this section.
}

We construct the HVA using the Kitaev couplings $J_x$, $J_y$ and $J_z$ in Eq.~\eqref{eq:HK}, and the magnetic fields  $h_x$, $h_y$ and $h_z$ in the Eq.~\eqref{eq:H}.
The HVA represented by the unitary matrix $U(\vec{\theta})$ has $L$ layers such that
\be
U(\vec{\theta}) = U_{\mathrm{HVA}}(\vec{\theta}_L) U_{\mathrm{HVA}}(\vec{\theta}_{L-1}) \cdots U_{\mathrm{HVA}}(\vec{\theta}_{1}).
\ee
Each layer $U_{\mathrm{HVA}}(\vec{\theta}_{\ell})$ consists of the exponentiation of the Kitaev couplings and the magnetic fields multiplied by the parameters $\vec{\theta}_{\ell}$ such that
\begin{align}
U_{\mathrm{HVA}}(\vec{\theta}_{\ell})
= & \, e^{-i \theta_{\ell;6}\sum_{i} Z_i} e^{-i \theta_{\ell;5}  \sum_{\text{z-bonds}} Z_i Z_j}
\nn\\
& \times e^{-i \theta_{\ell;4}\sum_{i} Y_i} e^{-i \theta_{\ell;3}\sum_{\text{y-bonds}}  Y_i Y_j}
\nn\\
& \times e^{-i \theta_{\ell;2} \sum_{i} X_i}e^{-i \theta_{\ell;1} \sum_{\text{x-bonds}} X_{i}X_{j}}.
\end{align}
The two-qubit gate count $N_b$ of each HVA layer scales linearly with the number $N$ of qubits. (For an infinite square-octagon lattice, $N_b = \frac{3}{2} N$.)
The Aspen-9 connectivity shown in Fig.~\ref{fig:lattice_geometry}(b) is natively the square-octagon lattice. One HVA layer can then be executed with $N_b \sim N$ CPHASE gates supported natively by Aspen-9 \cite{Reagor2018}.
If we execute the circuit on a QPU with a different two-dimensional connectivity, an overhead of $\mathcal{O}(N)$ SWAP gates will be required for implementing the HVA. Although this overhead does not change the overall gate complexity $\mathcal{O}(N)$ of the HVA implementation, it can still be quite demanding for NISQ devices.
Using the Aspen-9 QPU with the native connectivity thus provides a significant advantage in the near term.

There are six parameters for each HVA layer,  independent of the system size $N$. This makes the optimization of the HVA straightforward to analyze with increasing system size. We can make a rough estimate of how many times the cost function has to be evaluated based on the results on smaller lattices with a similar number of layers.
Nonetheless, if the ground state of a larger system exhibits longer-range entanglement, more layers will typically be needed to achieve the same degree of accuracy as in a smaller system.
Hence, the required number of layers and the total number of parameters depend on the system size, and usually can only be determined by testing.
This makes the prediction nontrivial especially when we consider small lattices in Fig.~\ref{fig:lattice_geometry} with noticeable finite-size effects.

Other than HVA, we consider the HEA which is widely used in NISQ applications \cite{Kandala2017,Havlicek2019,peters2021}. The general principle of the HEA is to construct circuit ansaetze using the native gates supported by the QPU, and hence different specific forms of the HEA are constructed targeting different QPUs. For Aspen-9, the native gate set consists of the XY gate \cite{Abrams2020}, the CPHASE and CZ gates \cite{Reagor2018} as well as the single-qubit gates $\mathrm{R}_{Z}(\theta)$ and $\mathrm{R}_{X}(k\pi/2)$ with $k=\pm 1,2$.
Moreover, the Quil programming language \cite{smith2016practical} and its accompanying optimizing compiler Quil-C~\cite{smith2020opensource} on the Rigetti stack \cite{Karalekas_2020} admit parametric compilation, which allows for the ansatz to be compiled only once, so that the numerical values of the ansatz parameters are only updated at runtime, without the need of incurring the compilation overhead at every step of the optimization process. This allows for faster execution times and feedback loops between the quantum and classical processors in the hybrid computation.

Roughly speaking, a HEA usually consists of layers of parameterized single-qubit rotations and two-qubit gates to create entanglement between qubits. Adopting this idea with respect to the Aspen-9 native gate set and its connectivity, we consider two ansaetze, namely the HEA-CZ using CZ gates and single-qubit rotation gates, and the HEA-XY using XY gates and single-qubit rotation gates. Here, we focus on the native two-qubit gates (CZ and XY) and slightly relax the native restriction on the single-qubit gates since the single-qubit gates have a much lower error rate than the two-qubit gates.
The unitary representation $U$ of an $N$-layer HEA is given by
\begin{align}
U(\vec{\theta}) = & \, U_{\mathrm{HEA}}(\vec{\theta}_L) \cdots
U_{\mathrm{HEA}}(\vec{\theta}_{1})  U_{0}(\vec{\theta}_{0}).
\end{align}
Here, the layer $U_{0}$ preparing each qubit in an arbitrary unentangled state is given by
\be
U_{0}(\vec{\theta}_{0}) = \prod_{i}  \mathrm{R}_{Z, i}(\theta_{0;1,i}) \mathrm{R}_{X, i}(\theta_{0;0,i}).
\ee
Note that an arbitrary single-qubit rotation is represented by the gate sequence $\mathrm{R}_Z \mathrm{R}_X \mathrm{R}_Z$. The rightmost $\mathrm{R}_Z$ can be omitted here since the qubits are initialized in the zero-state and $\mathrm{R}_{Z} \sket{0}$ just gives an irrelevant global phase.
The other layers $U_{\mathrm{HEA}}$ then entangle the qubits using the two-qubit gates.
For HEA-CZ, we have
\begin{align*}
U_{\mathrm{HEA}}(\vec{\theta}_{\ell}) = &
\prod_{j, k \in G} 
\mathrm{R}_{Z, j}(\theta_{\ell;4,(j,k)}) \mathrm{R}_{Z, k}(\theta_{\ell;3,(j,k)})
\nn\\
& \, \times \mathrm{R}_{X, j}(\theta_{\ell;2,(j,k)})  \mathrm{R}_{X, k}(\theta_{\ell;1,(j,k)})
\mathrm{CZ}_{j, k}
,
\end{align*}
where $G$ represents the hardware-native connectivity.
Similarly, for HEA-XY, we have
\begin{align*}
& U_{\mathrm{HEA}}(\vec{\theta}_{\ell})
\nn\\
= & \prod_{j, k \in G} 
\mathrm{R}_{Z, j}(\theta_{\ell;6,(j,k)}) \mathrm{R}_{Z, k}(\theta_{\ell;5,(j,k)})
\nn\\
& \, \times
\mathrm{R}_{X, j}(\theta_{\ell;4,(j,k)})  \mathrm{R}_{X, k}(\theta_{\ell;3,(j,k)})
\nn\\
& \, \times
\mathrm{R}_{Z, j}(\theta_{\ell;2,(j,k)}) \mathrm{R}_{Z, k}(\theta_{\ell;1,(j,k)})
\mathrm{XY}_{j, k}(\theta_{\ell;0,(j,k)})
	.
\end{align*}
Once again, the sequence $\mathrm{R}_Z \mathrm{R}_X \mathrm{R}_Z$ represents an arbitrary single-qubit rotation. For HEA-CZ, since $\mathrm{R}_z$ commutes with $\mathrm{CZ}$, the first $\mathrm{R}_z$ can be combined with the last $\mathrm{R}_z$ in the previous layer. $U_{\mathrm{HEA}}(\vec{\theta}_{\ell})$ thus requires two fewer single-qubit gates for HEA-CZ than for HEA-XY.

In the present case, with the native connectivity being the same as the lattice geometry, the number of two-qubit gates of the HEA scales as $N$, which is the same as for the HVA, as expected.
On the other hand, the number of parameters per layer also scales linearly with $N$, in contrast to the constant scaling of the HVA. The much larger number of parameters makes the HEA more expressive especially when a small number of layers is used. However, this also makes the HEA hard to scale up since optimizing a large number of parameters is challenging for non-convex cost functions.

\subsection{Optimizers}
\label{subsec:optimizers}

\begin{table}[t!]
	\caption{
		\textbf{Classical optimization algorithms and the different types of optimizers tested for the VQE algorithm.} These include gradient-free, gradient-based, and non-local optimizers, and a genetic algorithm. We use the SciPy \cite{SciPy2020} implementation for BFGS and Dual Annealing, a Fortran implementation developed by Powell \cite{Powell2009,BOBYQA2009} for BOBYQA, the pycma \cite{pycma2019} implementation for CMA-ES, and a python implementation developed by Gomez-Dans \cite{spsa2012} for SPSA.}
		\medskip
	\label{table:optimizers}
	\begin{tabular}{| c | c | c | c |}
		\hline
		Optimizer & Gradient-free  & Genetic & Local \\
		\hline
		BFGS & $\times$ & $\times$ & \checkmark\\
		\hline
		BOBYQA \cite{Powell2009} & \checkmark & $\times$ & \checkmark\\
		\hline
		CMA-ES \cite{Hansen2001} & \checkmark & \checkmark & $\times$ \\
		\hline
		Dual Annealing \cite{Xiang1997} & \checkmark & $\times$ & $\times$\\
		\hline
		SPSA \cite{Spall1992} & $\times$ & $\times$ & \checkmark\\
		\hline
	\end{tabular}
\end{table}

\begin{figure*}[t!]
	\centering
	\includegraphics[width=\textwidth]{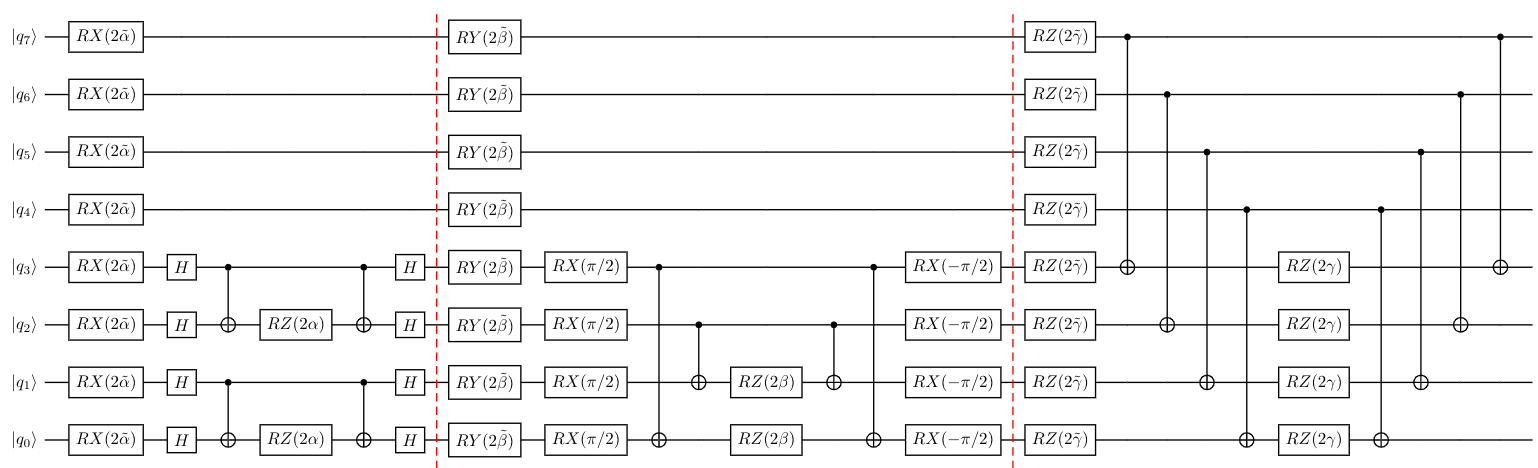}
	\caption{\textbf{HVA with one layer on eight qubits}.
		The Hamiltonian Variational Ansatz (HVA) with one layer on eight qubits splits into commuting blocks. The first block corresponds to the operation $e^{-i\tilde{\alpha} \sum_q X_q}$ $e^{-i\alpha \sum_{(i,j) \in X\text{-links}} X_i X_j}$, the second to $e^{-i\tilde{\beta} \sum_q Y_q}$ $e^{-i\beta \sum_{(i,j) \in Y\text{-links}} Y_i Y_j}$, and the third to $e^{-i\tilde{\gamma} \sum_q Z_q}$ $e^{-i\gamma \sum_{(i,j) \in Z\text{-links}} Z_i Z_j}$. For the circuit shown here, we used $X\text{-links} = \{(q_0, q_1), (q_2, q_3) \}$, $Y\text{-links} = \{(q_0, q_3), (q_1, q_2) \}$, and $Z\text{-links} = \{(q_0, q_4), (q_1, q_5), (q_2, q_6), (q_3, q_7) \}$.
		\edit{To execute this circuit, we map qubits $(q_0, q_1, q_2, q_3, q_4, q_5, q_6, q_7)$ to qubits $(12, 25, 26, 11, 13, 24, 27, 10)$, respectively, on Aspen-9 shown in \cref{fig:lattice_geometry}(b).}
	}
	\label{fig:hva-1-circuit}
\end{figure*}

\begin{figure*}[t!]
	\centering
	\includegraphics[width=\textwidth]{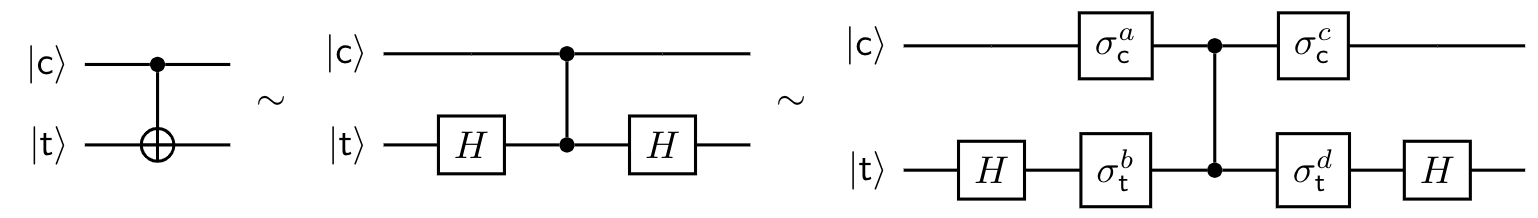}
	\caption{\textbf{Twirled CZ used to compile a CNOT}.
		Every use of the CNOT gate in Fig. \ref{fig:hva-1-circuit} is compiled to a native CZ with some single-qubit gates. Assuming that the major source of noise lies in the two-qubit gates, we apply random Pauli operators before every use of the CZ, and then apply another pair of Pauli operators in order to twirl the noise channel associated with the CZ into a stochastic Pauli channel. The relationship between the indices $a,b,c,d \in \{0,1,2,3\}$ (denoting respectively the Pauli operators $\{I,X,Y,Z\}$) is given in Eq.~\eqref{eq:twirl}. The subscript $c$ and $t$ on the Pauli matrices refer to "control" and "target" qubits. 
		All three circuits depicted above are logically equivalent, but generally lead to different noise characteristics.}
	\label{fig:twirled-cnot}
\end{figure*}

The cost function associated with VQE algorithms is in general non-convex with many local extrema. Similar to optimizing other classical non-convex cost functions, it is typically hard to decide on an efficient optimizer \emph{a priori} \cite{Rios2013}.
To make our benchmark less dependent on a specific choice of the optimizer, we will test the list of optimizers shown in Table~\ref{table:optimizers}.
This list consists of a few well-known gradient-free algorithms (BOBYQA, CMA-ES, dual annealing), gradient-based optimizers (SPSA, BFGS), non-local optimizers (CMA-ES, dual annealing), and a genetic algorithm (CMA-ES). Even though this is far from a comprehensive list, we can gain insight into how different types of optimizers work for our specific VQE cost function.

We compute the gradients needed for the gradient-based optimizers by finite-difference methods, which require an additional measurement of the cost function using the QPU for each gradient $\tfrac{\partial \langle E (\vec{\theta}) \rangle }{\partial \theta_j}$. This makes the cost of running these gradient-based optimizers larger than for the gradient-free algorithms, making them suboptimal given the limited clock rate of present-day QPUs. More efficient quantum algorithms to evaluate gradients have attracted great interest recently \cite{Romero_2018,Gilyen2019,Schuld19,Mitarai2020}. For instance, the parameter-shift rule allows the gradients to be evaluated with an analytic expression without introducing overhead in circuit depth or circuit width \cite{Schuld19}. The analytical evaluation will improve the performance of the optimization with noise. However, using the parameter-shift rule for the HVA ansatz requires a generator decomposition \cite{Izmaylov2021}, which makes the number of required expectation-value measurements scale with the system size $N$ (compared to constant scaling for finite-difference methods). Other algorithms usually require ancilla qubits and/or extra gate operations, and thus are challenging to implement on NISQ devices. Nonetheless, once QPUs with better gate fidelities and coherence properties are available, further studies will be required to revisit these results and strike a balance between quantum resource requirements and optimization efficiency.

\subsection{Determining energy gap by VQE}
\label{subsec:eneryg_gap}
The energy gap between the ground state and the first excited state is a useful quantity in probing the system's phase diagram.
VQE algorithms can obtain excited states by modifying the cost function with an additional term penalizing the overlap between the trial state and the lower-energy state obtained previously \cite{Higgott2019}. In particular, to determine the energy gap, we use the following cost function to determine the next excited state beyond the ground state,
\be
\label{app:cost_function_modified}
C(\vec{\theta}) =
\ssandwich{\psi(\vec{\theta})}{\psi(\vec{\theta})}{H}
+ p  |  \langle \psi(\vec{\theta}_{o;0}) | \psi(\vec{\theta}) \rangle |^2
,
\ee
where $p$ is the penalty weight and $\vec{\theta}_{o;0}$ are the optimal parameters corresponding to the ground state previously determined by VQE.
The overlap can be experimentally determined on QPUs by executing the circuit corresponding to $U^{\dagger}(\vec{\theta}_{o;0}) U(\vec{\theta}) \sket{0}$ and then measuring the population of the zero state $\sket{0}$.
The penalty weight $p$ is chosen to be larger than any relevant energy gaps, and we choose it to be 0.3 in our numerical experiment.

\subsection{Experiment and noise mitigation}
\label{subsecn:expt}

We run pre-optimized HVA circuits with one layer (HVA-1) on four- and eight-qubit sub-lattices of Rigetti's Aspen-9 QPU.
The variational optimization of the HVA is first performed on a classical simulator, and the circuits are then run on the QPU for the optimal choice of parameters. While this approach circumvents the hybrid quantum/classical computation, it nevertheless serves as a benchmark for how well the quantum processor performs in proof-of-concept experiments.  For comparison,  recent state of the art results for many body Hamiltonian simulations have opted for either purely classical optimization instead of a VQE step~\cite{huggins2021} or running their algorithms on smaller numbers of qubits~\cite{clustervqe,klymko2021}. 

\begin{table*}[t!]
	\begin{center}
		\caption{\textbf{Optimizer performance with statevector simulator}. We test the optimizers with eight qubits and 16 qubits using a four-layer HVA and the (GL+$h$) parameter set. We employ a multiple-initial-value strategy to BFGS and BOBYQA to avoid the local optimizers being trapped at a bad local extreme. The lowest energy is reported for the multiple-initial-values strategy. While the genetic algorithm (CMA-ES) is less susceptible to local extrema, a multiple-initial-values strategy with a smaller iteration cutoff also improves its performance. SPSA and dual annealing converge slowly and reach the maximum iteration (cutoff). Note that the error in energy is the difference between the optimized energy and the ground-state energy computed by ED.
		}
		\label{table:optimizer_performance}
		\begin{tabular}{ |c|c|c|c| } 
			\hline
			Qubits  & Optimizer & Error in energy  & Cost function  evaluations \\
			\hline
			8  & BFGS, 501 initial values & 0.00094  &  mean: 5352, max: 10865\\
			\hline
			16  & BFGS, 501 initial values & 0.02672 & mean: 5569, max: 10186\\
			\hline
			8  & BOBYQA, 501 initial values & 0.00045  &  mean: 1099, max: 1920\\
			\hline
			16  & BOBYQA, 501 initial values & 0.01744  & mean: 1255, max: 3049 \\
			\hline
			8  & CMA-ES & 0.00015   & 43290 \\
			\hline
			16  & CMA-ES & 0.04036   & 49335 \\
			\hline
			8  & CMA-ES, 80 initial values & 0.00005  &  mean: 20528, max: 83590\\
			\hline
			16  &  CMA-ES, 80 initial values  & 0.01327 & mean: 27504, max: 73138\\
			\hline
			8  & Dual annealing & 0.00252   & 100000 (cutoff) \\
			\hline
			16  & Dual annealing & 0.04634   & 100106 (cutoff) \\
			\hline
			8  & SPSA & 0.04500   & 100000 (cutoff) \\
			\hline
			16  & SPSA & 0.08145   & 100000 (cutoff) \\
			\hline
		\end{tabular}
	\end{center}
\end{table*}

We perform digital zero-noise extrapolation (ZNE) \cite{PhysRevX.7.021050,PhysRevLett.119.180509,9259940} by first exaggerating the errors in the circuit by replacing a subset of CZs by an odd multiple of them. We choose such subsets to increase the total CZ count in the circuit by a factor of $\lambda$.
For each scale factor $\lambda$, we run 25-100 (four qubits) and 100 (eight qubits) different circuit implementations for 1000 shots each.
For the four-qubit case, we pick $\lambda = 1, 1.3, 1.6, 2, 2.3.$
The bare HVA-1 circuit, with six CZs, corresponds to $\lambda = 1.0$. To obtain $\lambda > 1.0$, we note that the exponentials of each of the two-local terms ($XX$, $YY$ and $ZZ$) are compiled to two CZs, along with a few one-qubit gates. For $\lambda = 1.3$, we replace the first of the two CZs compiling the exponential of the $XX$ term with three CZs, then again for the $YY$ and $ZZ$ terms, then take the average of all three types of circuits to get an estimated expectation value of the cost Hamiltonian. For $\lambda = 1.6$, we replace both the CZs with three CZs each, once for each of the three two-local terms. For $\lambda = 2.0$, we replace both CZs in the $XX$ term, and the first of the two CZs in the $YY$ term  with three CZs each for one experiment, and for another we replace both CZs in the $ZZ$ term and the second of the two CZs in the $YY$ term with three CZs each. Finally, for $\lambda = 2.3$, we perform three experiments, in which we replace the CZs in the $XX$ and $YY$, $XX$ and $ZZ$, or $YY$ and $ZZ$ exponentials with three CZs each.

For the eight-qubit case, we choose $\lambda=1, 1.5, 2, 3, 5$.
For $\lambda = 1$, we have a total of 16 CZs, two for each of the edges. For $\lambda = 1.5$, we run 25 different circuits for each of the scenarios where we replace (a) every CZ in the $XX$-links with three CZs, (b) every CZ in the $YY$-links with three CZs, (c) every CZ in two of the four $ZZ$-links with three CZs, and (d) every CZ in the remaining two of the four $ZZ$-links with three CZs. For $\lambda = 2$, we run 50 different circuits for each of the scenarios where we replace (i) every CZ in the $XX$- and $YY$-links with three CZs, and (ii) every CZ in the $ZZ$-links with three CZs. For $\lambda = 3$ ($\lambda = 5$), we simply take 100 different circuits where we replace every CZ in the circuit with three (five) CZs.

For each of the circuit implementations in each scale factor family, we estimate the cost function \eqref{eq:cost}
\edit{
by measuring the expectation values of all Pauli strings given in \cref{eq:H}.
Each of these expectation values associated with $N$ qubits is determined experimentally by applying suitable single-qubit gates to transform any Pauli $X$ or Pauli $Y$ to Pauli $Z$ followed by a projective measurement over the computational basis. The measured probability distributions are subjected to readout errors which can be mitigated as follows.
The readout errors can be modeled by classical Markovian process \cite{Geller2020} described by the equation
\be
p_{\mathrm{read}} = R \, p_{\mathrm{prepared}}.
\ee
Here, $R$ is the $2^N \times 2^N$ confusion matrix that relates the probability distribution $p_{\mathrm{prepared}}$ of the prepared state without readout error and the probability distribution $p_{\mathrm{read}}$ experimentally measured with readout error. Each element of $R$ can be interpreted as the conditional probability of measuring the bitstring $x_\mathrm{read}$ given that the bitstring $x_{\mathrm{prepared}}$ is prepared.
In our experiment, $R$ is estimated by preparing a bitstring, then computing the fraction of each of $2^N$ bitstrings from 10,000 shots to estimate the conditional probabilities, then repeating this procedure for all $2^N$ bitstrings.
By inverting $R$, we obtain a readout error mitigated estimate of the probability of outcomes of all bitstrings using
\be
p_{\mathrm{prepared}} = R^{-1} \, p_{\mathrm{read}}.
\ee}
In turn, these lead to readout-error-mitigated estimates of the cost function~\eqref{eq:cost} in each of the 100 circuits for any given family associated with a scale factor. We then average these expectation values over all 100 circuits to produce a single value representing the family of circuits associated with a given scale factor. These representative values are then plotted against the scale factors, and a polynomial fit is found to extrapolate to the $\lambda = 0$ (zero noise) limit to obtain an estimate of the cost function mitigated for both gate and readout errors.

\begin{figure*}[t!]
	\centering
	\includegraphics[width=\textwidth]{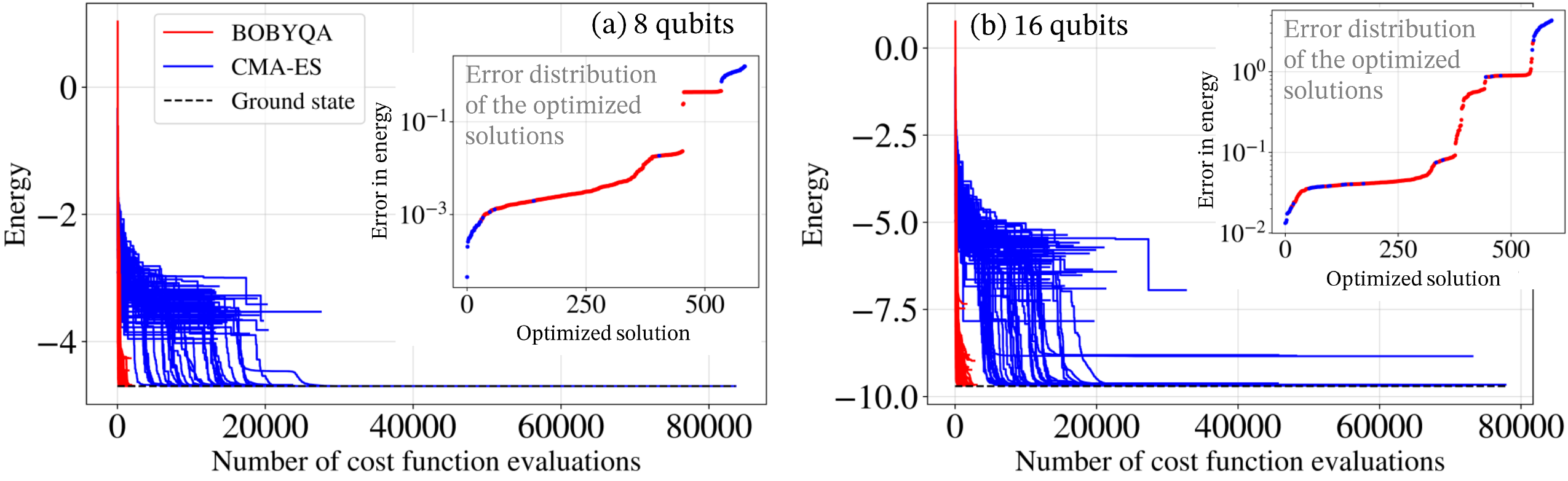}
	\caption{\textbf{Noiseless optimization with BOBYQA and CMA-ES}.
		The cost function associated with the four-layer HVA and the parameter set (GL + $h$) is optimized by BOBYQA (red) with 501 random initial values and CMA-ES (blue) with 80 random initial values.
		For both (a) eight qubits and (b) 16 qubits, BOBYQA converges faster than CMA-ES. The error distribution of the optimized solutions (see insets \edit{showing the error in energy associated with each solution optimized by BOBYQA and CMA-ES}) shows that the solutions of CMA-ES give overall better results but are also less consistent. The mixed usage of the two optimizers provides a complementary cross-check of their results.}
	\label{fig:optimization_track}
\end{figure*}

	For the eight-qubit case, we also perform randomized compilation \cite{randomized-compilation} to twirl the physical error channels into stochastic errors to improve the result.
In the HVA, the exponential of every two-body term can be compiled using two native CZs. Before each CZ gate, we apply random Pauli operators $\sigma_{\mathsf c}^{a} \sigma_{\mathsf t}^{b}$, where $a,b=0,\dots,3$ and the subscripts $\mathsf c$ and $\mathsf t$ denote the control and target qubits, respectively. After applying the CZ unitary $\Lambda = \vert 0 \rangle \langle 0 \vert_{\mathsf c} \otimes \mathbb{I}_{\mathsf t} + \vert 1 \rangle \langle 1 \vert_{\mathsf c} \otimes {Z}_{\mathsf t}$~\footnote{Note that the same unitary can also be written $\mathbb{I}_{\mathsf{c}} \otimes \vert 0 \rangle \langle 0_{\mathsf{t}} + \mathbb{Z}_{\mathsf{c}} \otimes \vert 1 \rangle \langle 1 \vert_{\mathsf{t}}$, so that the CZ gate remains invariant if we swap the labels $\mathsf{c}$ and $\mathsf{t}$. Therefore, it doesn't matter how we assign the labels control/target to the two qubits participating in this interaction.}, we apply Pauli operators $\sigma_{\mathsf c}^{c} \sigma_{\mathsf t}^{d}$ such that $\sigma_{\mathsf c}^{c} \sigma_{\mathsf t}^{d} = \Lambda \sigma_{\mathsf c}^{a} \sigma_{\mathsf t}^{b} \Lambda^{\dagger}$.
Following Ref.\ \cite{PhysRevX.7.021050}, this is given by 
\begin{align}
	\label{eq:twirl}
	\begin{split}
		c &= a + b(3-b)(3-2a)/2,\\
		d &= b + a(3-a)(3-2b)/2.
	\end{split}
\end{align}
This noise tailoring makes the errors more well behaved and better suited for error mitigation techniques such as ZNE~\cite{PhysRevX.7.021050, Schultz-PRA-2022,Cai-EM_review-arXiv-2022,Russo-arXiv-2022}. A schematic circuit for the HVA with one layer on an eight-qubit sub-lattice is shown in Fig. \ref{fig:hva-1-circuit}. The twirling operation is depicted in Fig. \ref{fig:twirled-cnot}.

	For twirled experiments we take point-estimates of the expectation values for many circuit implementations for each scale factor and feed them into a Bayesian linear regression model implemented in Turing.jl \cite{ge2018t} with Gaussian priors for slope and intercept and a truncated-at-zero Gaussian prior for the variance, and we run Markov Chain Monte Carlo (MCMC) until well converged. Hyperparameters of the priors were all set to mean zero and standard deviation 1000 (i.e.\ no prior information or preference over the reasonable range of the parameters), but these hyperparameters had negligible impact on the posterior distribution so long as they were not set to effectively exclude the region of the observed data. MCMC was run with a No U-Turn Sampler \cite{hoffman2014no} for $10^4$ iterations, we exclude the first 3000 to exclude any warmup period, and for all runs we had an effective remaining sample size $>1000$.
	For untwirled experiments, we follow the same basic procedure but incorporate shot noise for each circuit implementation by performing a Bayesian bootstrap \cite{rubin1981bayesian} over the shot data to produce an estimate and uncertainty for the expectation value from each circuit implementation. We then feed those estimates and uncertainties into the Bayesian linear regression defined above.	

\section{Results}
\label{sec:results}
\subsection{Statevector simulations}

\begin{figure}[t!]
	\centering
	\includegraphics[width=\linewidth]{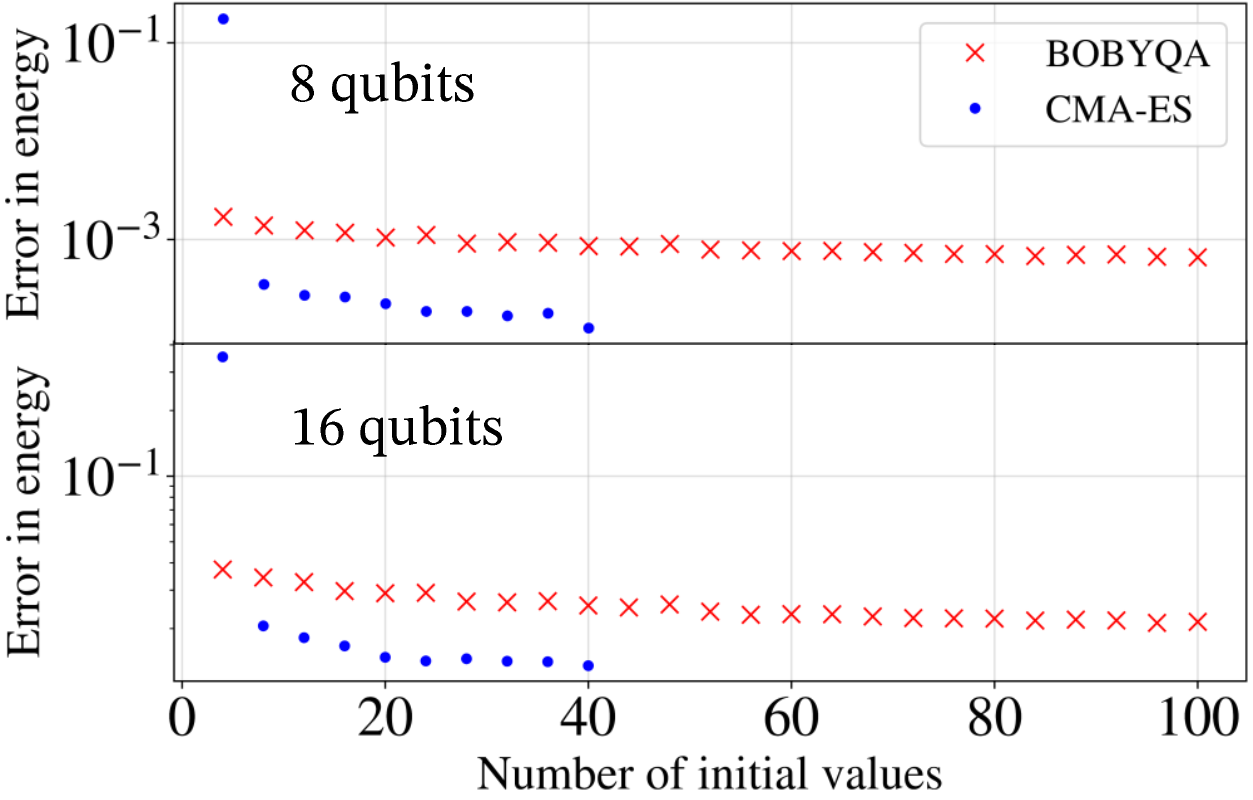}
	\caption{
		\textbf{Error as a function of the number of initial parameter values for the BOBYQA and CMA-ES optimizers}.
		The error decreases with increasing number of initial values. CMA-ES converges faster compared to BOBYQA at the price of a much higher number of cost function evaluations per initial value. The performance of CMA-ES exceeds that of BOBYQA after a threshold number of initial values. The performance of both methods improves gradually beyond that threshold.
		We resample the error distribution shown in \cref{fig:optimization_track} to obtain this result. Each data point in this plot is obtained by averaging the minimum energies of 50 different random choices of initial values. The statistical error (not shown) is negligible except for the leftmost CMA-ES data point, and this is consistent with the fact that CMA-ES is unreliable with a small number of initial values.
	}
	\label{fig:optimization_resampling}
\end{figure}

\begin{figure*}[t!]
	\centering
	\includegraphics[width=0.9\textwidth]{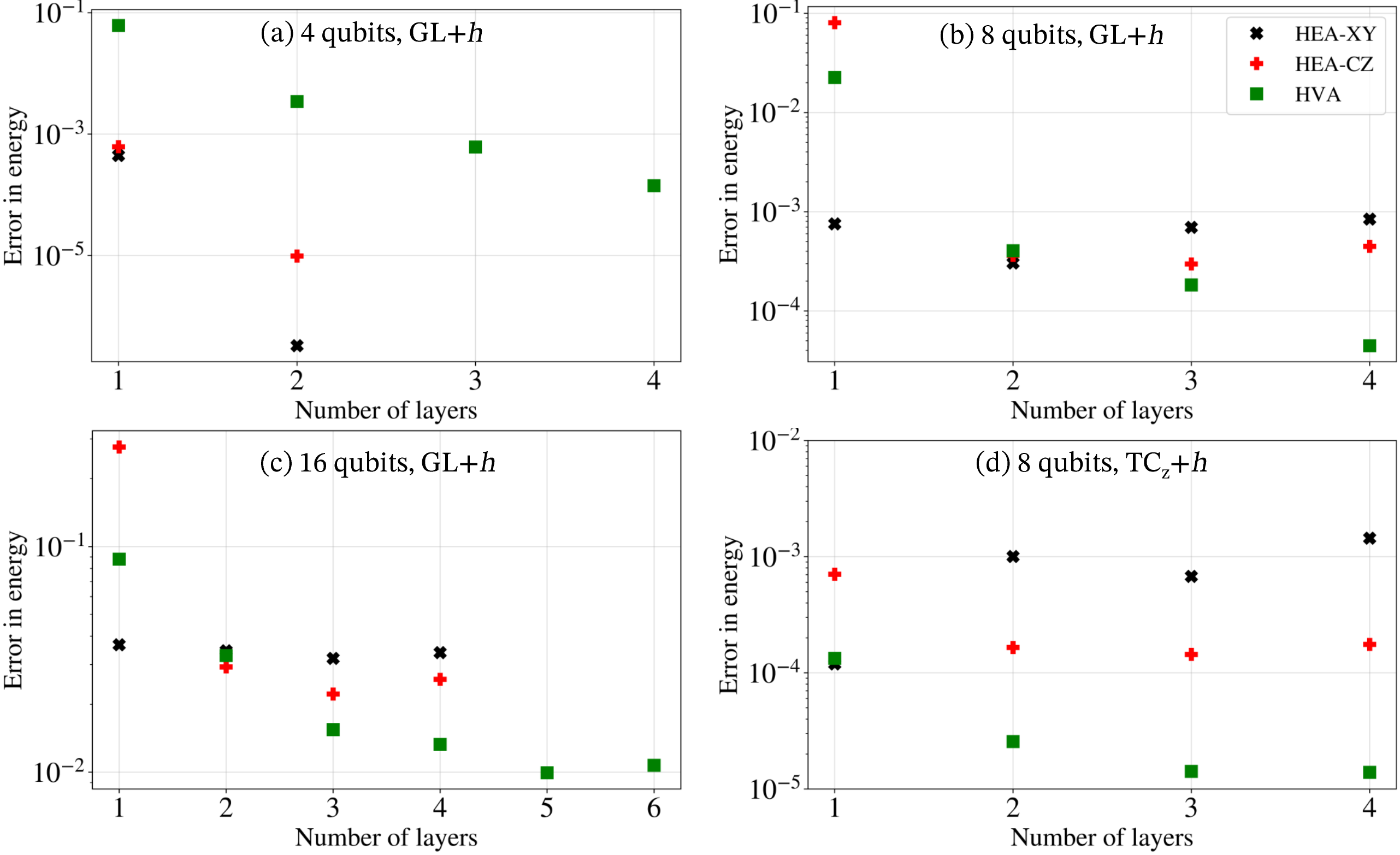}
	\caption{\textbf{Ansatz benchmark by statevector simulation.}
		The three different ansatzes are optimized with respect to three different setups, (a) four-qubit lattice with (GL +$h$), (b) eight-qubit lattice with (GL +$h$), (c) 16-qubit lattice with (GL +$h$) and (d) eight-qubit lattice with (TC$_z$+$h$). The HEA-CZ (\edit{red pluses}) and the HEA-XY (black crosses) show a bigger error in energy compared to that of the HVA (\edit{green squares}) except for the four-qubit case. This performance difference is mainly due to the difficulty in optimizing the two HEAs with a large number of parameters, as suggested by the minimal improvement of the error when increasing the number of layers of the HEAs. }
	\label{fig:vqe_simulation}
\end{figure*}

We test the optimizers listed in Table~\ref{table:optimizers} using the cost function~\eqref{eq:cost} associated with the four-layer HVA for the eight-qubit and 16-qubit lattices shown in Fig.~\ref{fig:lattice_geometry} and the parameter choice (GL + $h$) in Table~\ref{tab:model_params}. This cost function has 24 parameters to be optimized. The results using a statevector simulator are summarized in Table~\ref{table:optimizer_performance}.

BFGS and BOBYQA are susceptible to local extrema and their performance highly depends on the initial choice of the parameters. We employ a multiple-initial-condition strategy and randomly pick 501 initial parameter values to start with. In general, the number of initial values required for a good result will increase with the number of parameters of the cost function. BOBYQA gives a smaller error in energy compared to BFGS in both eight-qubit and 16-qubit tests, and BOBYQA also converges much faster than BFGS, which requires the costly computation of gradients by finite-difference methods.
The other local optimizer, SPSA, is less susceptible to local extrema due to its stochastic nature. However, its convergence is slow and will require many iterations in order to achieve good accuracy.
Among the local optimizers, BOBYQA with multiple initial values gives the best performance in terms of both the accuracy of the result and the speed of convergence.

CMA-ES and dual annealing are non-local optimizers that are able to escape from local extrema. Table~\ref{table:optimizer_performance} shows that both optimizers converge toward a minimum with energy close to the exact value even starting from just one initial value. CMA-ES consistently converges faster among the two.
Obtaining the best performance from CMA-ES requires fine tuning of the optimizer meta-parameters, which depends on the neighborhood of the initial value and is challenging to determine \emph{a priori}. To avoid becoming overly sensitive to the choice of meta-parameters, we also employ a multiple-initial-value strategy with 80 random initial values.
For our VQE application, the CMA-ES optimization with multiple initial values consistently shows improvement over that with a single initial value.

To benchmark different ansaetze, we employ both BOBYQA with 501 random initial values and CMA-ES with 80 random initial values and report the better result from the two methods. Different ansaetze (and different layer numbers) result in vastly different cost-function landscapes and numbers of parameters. A mixed usage of different optimization algorithms, which is widely used in black-box optimization, helps us better navigate through the vast varieties of cost functions.
In Fig.~\ref{fig:optimization_track}, we study the noiseless optimization for each optimizer with different initial values. For both eight-qubit and 16-qubit cases, CMA-ES needs more cost function evaluations to converge, as expected for a non-local optimizer.
As indicated by the error distribution of the optimized solutions shown in the insets of Fig.~\ref{fig:optimization_track}, some optimization runs with CMA-ES give the best result but others can be far off. The runs also converge to different optimized parameters with the two different optimizers and also different initial values, revealing a complex energy landscape with many local minima.
This makes the inclusion of the BOBYQA results helpful, since it allows us to crosscheck the optimizations.
A similar conclusion can be made by resampling the distribution shown in the insets of Fig.~\ref{fig:optimization_track} to investigate the error as a function of the number of initial values of optimization. As shown in \cref{fig:optimization_resampling}, the CMA-ES optimizer gives a better result than BOBYQA after a sufficient number of initial values are used. Hence, BOBYQA serves as a useful cross-check for an insufficient number of initial values.
We expect that VQE optimization carried out on QPUs will also benefit from this approach of mixed optimizers with multiple initial values, especially when the QPUs' clock rate becomes sufficiently high to support a large number of cost function evaluations.

We apply this mixed optimization approach to optimize the three different ansaetze (HVA, HEA-CZ and HEA-XY) with different numbers of layers using a statevector simulator. To make our observation more representative, we run the optimization test with three different setups, namely, four-qubit (GL+$h$), eight-qubit (GL+$h$), 16-qubit (GL+$h$) and eight-qubit (TC$_z$+$h$) (see Table~\ref{tab:model_params} for the parameters values).
Except for the simple four-qubit case, the results in Fig.~\ref{fig:vqe_simulation} show that the HVA in general outperforms the other two HEAs when a deeper ansatz is used to reduce the error in energy.
The main reason for the HEAs' worse performance is due to the extreme difficulty in the optimization, which is directly related to the barren plateau phenomenon \cite{McClean2018}. The number of parameters increases rapidly for HEAs; for example, the four-layer HEA-XY for 16 qubits has 536 parameters. This makes it excessively challenging to scale up the VQE with HEAs to larger circuit depths and system sizes.
This is also revealed by the fact that two HEAs show nearly no improvement with increasing number of layers in \cref{fig:vqe_simulation}.
Hence, the VQE using HVA will be a more promising approach toward quantum advantage for the square-octagon-lattice Kitaev model.

\begin{figure}[t!]
	\centering
	\includegraphics[width=\linewidth]{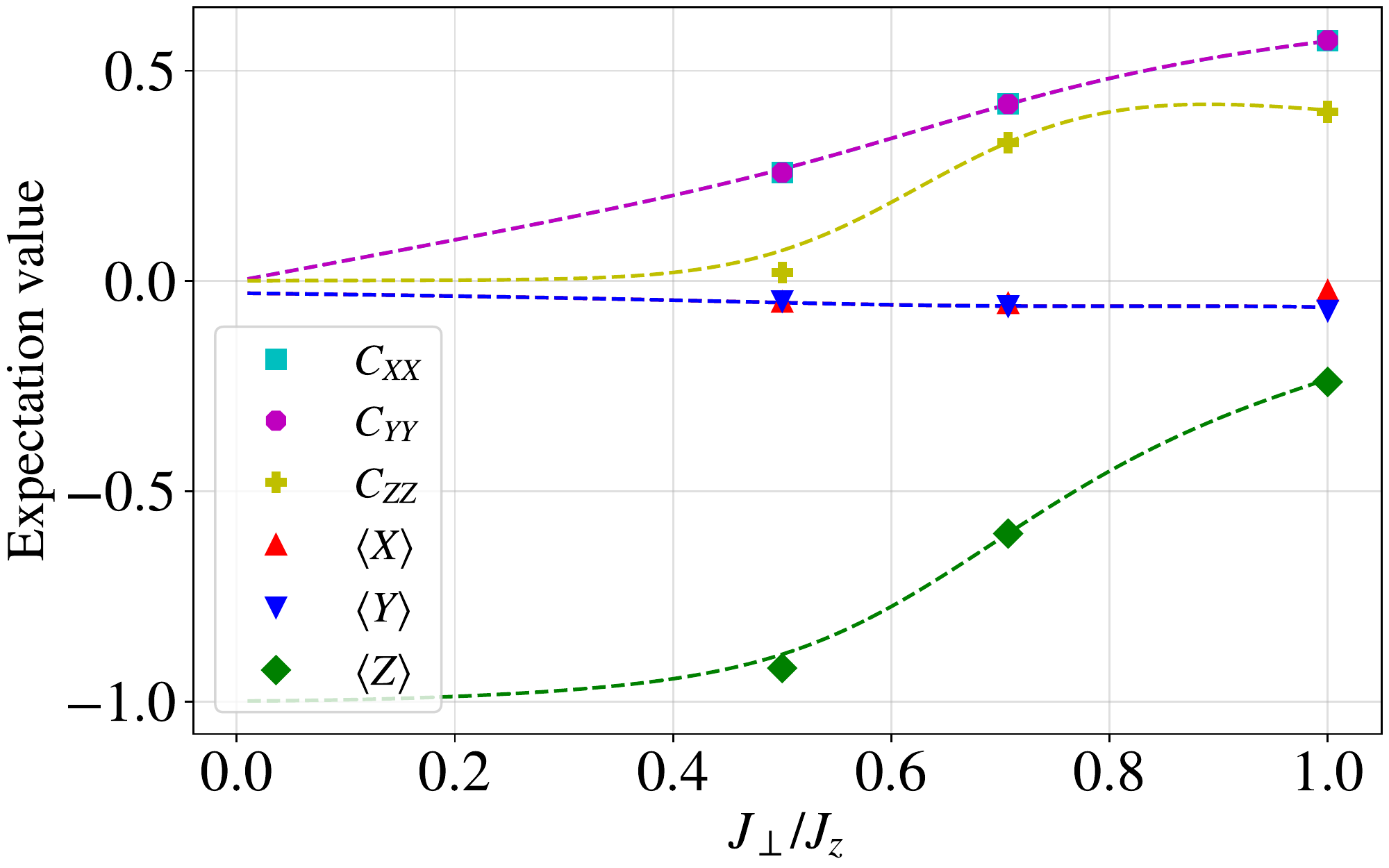}
	\caption{
			\textbf{Expectation values determined by VQE for an eight-qubit periodic lattice}.
			The VQE results (data points) are consistent with those obtained from ED (dashed lines). Here, the single-qubit expectation values $\langle X \rangle$, $\langle Y \rangle$ and $\langle Z \rangle$ are averaged over all qubits, and the correlators $C_{\alpha \alpha} = \langle \alpha_j  \alpha_k \rangle - \langle \alpha_j \rangle \langle \alpha_k \rangle$ ($\alpha = X, Y, Z$) are averaged over the qubit pairs for which $J_{\alpha}\neq0$. The VQE results are determined using an eight-layer HVA ansatz with the optimization using 480 initial values for CMA-ES and 3001 initial values for BOBYQA, and the expectation values are calculated using the optimized parameters corresponding to the lowest energy among all optimized solutions. We use the parameters $J_z = 1$ and $h_x = h_y = h_z = 0.05 / \sqrt{3}$.
	}
	\label{fig:expectation_values}
\end{figure}

\begin{figure}[t!]
	\centering
	\includegraphics[width=\linewidth]{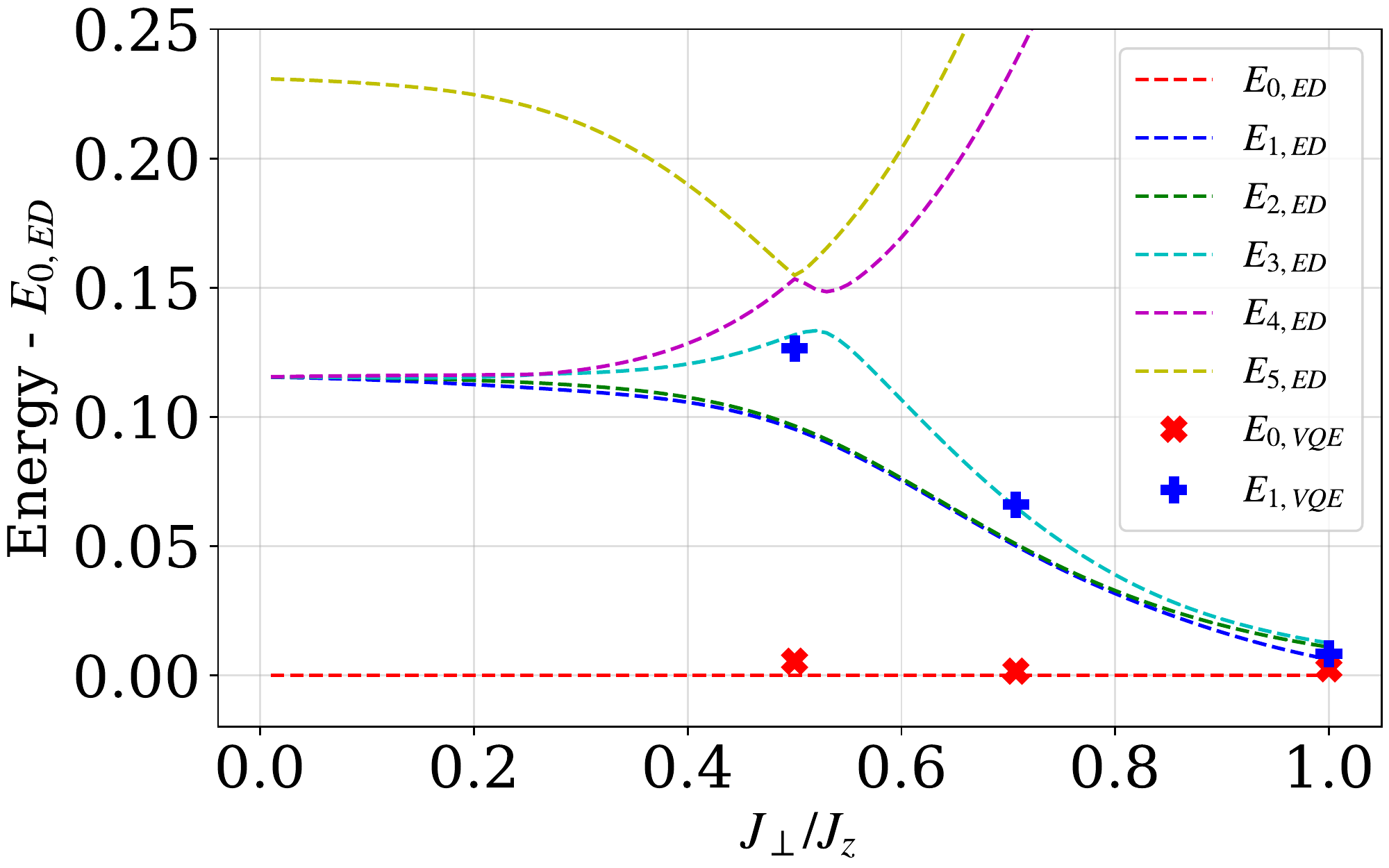}
	\caption{
			\textbf{Excited-state and ground-state energies determined by VQE for an eight-qubit periodic lattice}.
			The ground-state energies $E_{0, \text{VQE}}$ (red crosses) and the first excited-state energies $E_{1, \text{VQE}}$ (blue pluses) determined by VQE are compared to the low-lying energy spectrum (dashed lines) determined by ED. VQE correctly predicts the qualitative result of the closing of the energy gap near $J_\perp = J_z$. However, the excited state energies determined by VQE are closest to the third excited state instead of the nearly degenerate first and second excited states.
		We use the parameters $J_z = 1$ and $h_x = h_y = h_z = 0.05 / \sqrt{3}$.
	}
	\label{fig:excited_states}
\end{figure}

The optimization approach and the HVA ansatz can be used to probe the phase diagram of the model. We demonstrate this potential by determining the single-spin expectation values and static spin-spin correlators in \cref{fig:expectation_values} for an eight-qubit lattice with periodic boundary conditions as shown in \cref{fig:lattice_geometry}(f).
We choose this lattice structure and boundary condition so that the system's behavior is less susceptible to boundary effects. Note that the lattice contains four $x$, $y$, and $z$ bonds each. To adapt to this different setup, we have to increase the number of layers of the HVA ansatz to 8 to achieve reasonable accuracy. To characterize the system's properties as a function of Hamiltonian parameters, we measure the averages of the single-site expectation values $\langle X \rangle$, $\langle Y \rangle$, and $\langle Z \rangle$ over all sites, as well as the averages of the correlators
$C_{\alpha \alpha} = \langle \alpha_j  \alpha_k \rangle - \langle \alpha_j \rangle \langle \alpha_k \rangle$ ($\alpha = X, Y, Z$) over all pairs of sites for which $J_\alpha\neq 0$. While this small system does not exhibit the exact same phase transitions shown in \cref{fig:phase-diagram} for large lattices in the thermodynamic limit, the same set of measurements can be used to study the phase diagram of larger systems. 
Increasing $J_{\perp}$ with a fixed $h_{[111]}$, the eight-qubit lattice shows an increasing average value of $\langle Z \rangle$ while $\langle X \rangle$ and $\langle Y \rangle$ remain close to zero. Note that the asymmetry between $\langle Z \rangle$ and $\langle X\rangle$ or $\langle Y \rangle$ at $J_{\perp} / J_z = 1$ is due to the asymmetry in the lattice geometry. Meanwhile the average correlators $C_{\alpha \alpha}$ all increase gradually from zero as $J_\perp$ grows. This behavior is captured by the VQE optimization results (data points) as compared to the simulation using ED (dashed lines). 
One can observe a small deviation between the VQE and ED results for $C_{ZZ}$ at $J_{\perp}/J_z = 0.5$. This could be improved by increasing the number of layers used with the trade off of higher resource requirements in the optimization.

The phase diagram can also be probed by searching for parameter values where the energy gap between the ground and first excited states closes.
By adding the overlap with the ground state to the VQE cost function as discussed in \cref{subsec:eneryg_gap}, we determined the excited-state energies using the same HVA ansatz and optimization approach.
For the eight-qubit periodic lattice, we determine the optimal parameters $\vec{\theta}_{o;0}$ corresponding to the ground states for different values of $J_{\perp}$ using the eight-layer HVA ansatz as discussed in the previous paragraph.
For excited states, we optimize the modified cost function in \cref{app:cost_function_modified} with the same eight-layer HVA ansatz. For the optimization, we use CMA-ES with 960 ($J_{\perp} = 0.5,1/\sqrt{2}$) and 480 ($J_{\perp} = 1$) random initial values, and BOBYQA with 6001 ($J_{\perp} = 0.5, 1/\sqrt{2}$) and 3001 ($J_{\perp} = 1$) random initial values. We double the number of initial values for $J_{\perp} = 0.5, 1/\sqrt{2}$ as an attempt to improve the accuracy of the excited-state energies determined by the optimization.

The ground- and excited-state energies determined by VQE are shown in \cref{fig:excited_states}. To better visualize the results, we offset the energy shown in the figure by the ground-state energy $E_{0, ED}$ determined using ED. We also plot the ED result ($E_{j, ED}$ where $j=0,1,\cdots, 5$) for up to the fifth excited state for comparison. As expected, the VQE results (red crosses) for the ground-state energy are consistent with the ED result (red dashed line) with a small energy error.
On the other hand, the excited-state energies (blue pluses) determined by VQE are closest to those of the third excited state instead of the nearly degenerate first and second excited states (except for $J_{\perp}=1$ where the first three excited states are close to each other). This means that while the VQE result still qualitatively predicts the closing of the energy gap, it does not give the correct energy gap between the ground state and the first excited state.
We tried multiple values of $p$ and observed a similar difficulty in getting to the lower-energy excited states. Future study of alternative ansatz better tailored to the excited states will be needed to improve the accuracy.

\begin{figure}[t!]
	\centering
	\includegraphics[width=\linewidth]{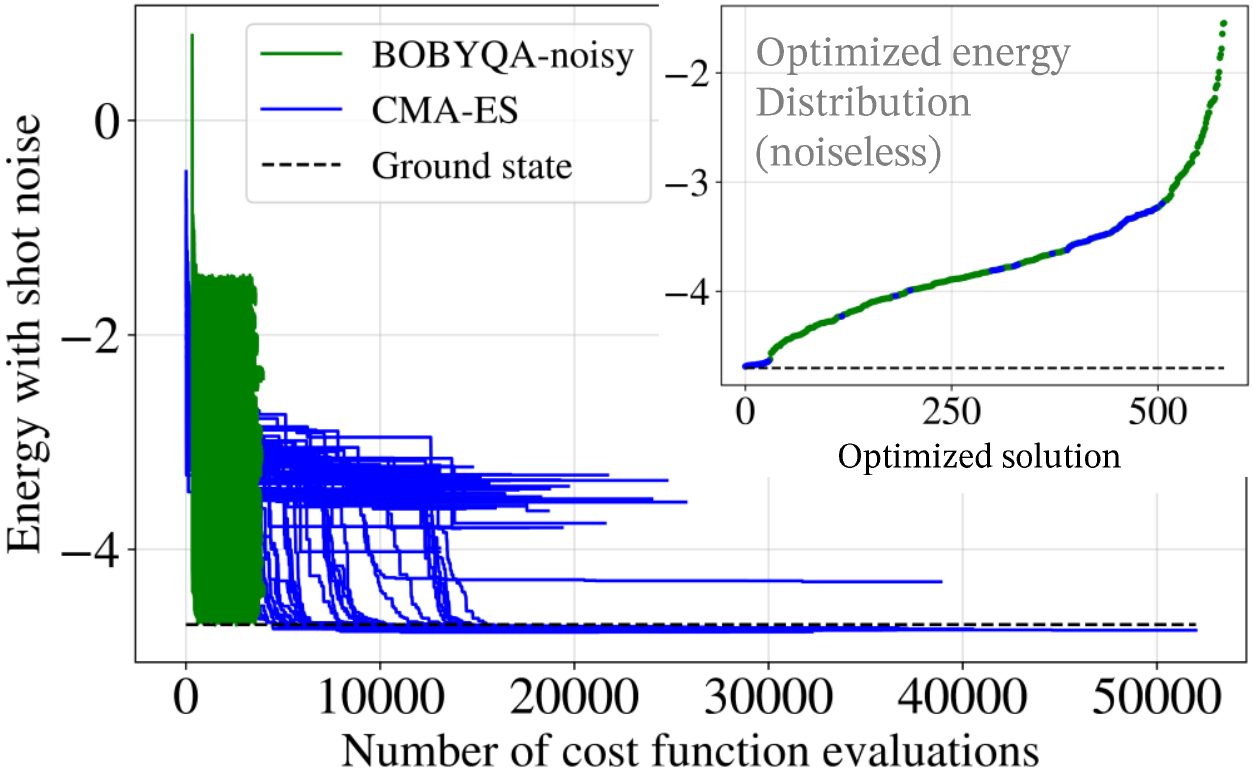}
	\caption{\textbf{eight-qubit Optimization with shot noise using BOBYQA-noisy and CMA-ES}.
		The cost function associated with the four-layer HVA and the parameter set (GL + $h$) is optimized by BOBYQA-noisy (green) with 501 random initial values and CMA-ES (blue) with 80 random initial values.
		Similar to the noiseless optimization, BOBYQA-noisy converges faster than CMA-ES but the solutions of CMA-ES give overall better results as shown by the optimized energy distribution (see insets).
	}
	\label{fig:optimization_shotnoise_track}
\end{figure}

\begin{table*}[t!]
	\begin{center}
		\caption{\textbf{Optimizer performance with shot noise for eight qubits}.
			We test the optimizers with 8000 measurement shots and eight qubits using a four-layer HVA and the (GL+$h$) parameter set similar to the noiseless case.
			A modified version of BOBYQA, BOBYQA-noisy, is introduced to better handle the noisy cost function.
			Using the optimized parameters obtained by minimizing the noisy cost function, we determine the energy $E_{\mathrm{noisy}}$ measured with shot noise and its noiseless counterpart $E_{\mathrm{noiseless}}$ obtained using a noiseless statevector simulator.
			The standard deviation of 100 energy evaluations with 8000 shots is about $0.02$ for the optimized parameters obtained by these optimizers.
			Note that $E_{\mathrm{noisy}}$ can be lower than $E_0$ due to the shot noise.
		}
		\label{table:optimizer_performance_shotnoise}
		\begin{tabular}{ |c|c|c|c| } 
			\hline
			Optimizer &   $E_{\mathrm{noiseless}} - E_0$  &  $E_{\mathrm{noisy}} - E_0$  & Cost function evaluations\\
			\hline
			BFGS, 501 initial values  & 0.45069 & 0.42052 &  mean: 747, max: 1994 \\
			\hline
			BOBYQA, 501 initial values & 0.27485  & 0.21843 & mean: 471, max: 610\\
			\hline
			BOBYQA-noisy, 501 initial values & 0.07989 & -0.00453 &  mean: 3532, max: 4004\\
			\hline
			CMA-ES & 0.02416 & -0.06462 & 37570 \\
			\hline
			CMA-ES, 80 initial values & 0.01610 & -0.07125 &  mean: 21042, max: 52000 \\
			\hline
			Dual annealing & 0.04534 & -0.01631  & 60101 \\
			\hline
			SPSA & 0.00612 & 0.00879 & 100000 (cutoff) \\
			\hline
		\end{tabular}
	\end{center}
\end{table*}

\subsection{Simulations with shot noise}

It is crucial to take the shot noise into consideration when running variational algorithms on QPUs. If too many measurement shots are requested, the long runtime may be beyond the QPU capacity. If too few shots are requested, the stronger noise may induce undesirable barren plateaus to the cost function landscape \cite{wang2021noiseinduced}.
Investigating optimization strategies that tolerate a small number of shots is important for practical applications of the variational algorithms \cite{gu2021}.

In this subsection, we evaluate the energy in the simulations with a finite number of shots and then test the optimizers listed in Table~\ref{table:optimizers} once again with the noisy cost function. This allows us to examine the robustness of our optimization approach in the presence of the shot noise.
We run the test with 8000 shots using the cost function associated with the four-layer HVA for the eight-qubit lattice and the parameter choice (GL + $h$) to facilitate the comparison with the noiseless results. The results are summarized in Table~\ref{table:optimizer_performance_shotnoise}.
To make the effect of the shot noise clear, we determine the energy $E_{\mathrm{noisy}}$ measured with shot noise and its noiseless counterpart $E_{\mathrm{noiseless}}$ obtained using a statevector simulator for the optimized parameters obtained by minimizing the noisy cost function. The error $(E_{\mathrm{noiseless}} - E_0)$ from the exact ground-state energy $E_0$ shows the performance of the VQE since this quantity is directly related to how close the optimized state is to the ground state. The measured deviation $(E_{\mathrm{noisy}} - E_0)$ gives us additional information regarding the performance of the optimizers in the presence of the shot noise.

BFGS shows significantly worse performance compared to the noiseless case. The errors introduced by the shot noise are amplified by the small finite step size used by finite-difference methods to compute gradients. The amplified errors make the BFGS much more likely to be trapped. Using analytical methods to determine the gradients directly through measurements \cite{Romero_2018,Mitarai2020} could avoid the amplification and would be important when implementing BFGS (and similar gradient-based methods) on QPUs.

The performance of BOBYQA is also worse than that in the noiseless case.
This observation is consistent with previous results indicating that BOBYQA typically requires further modifications to optimize the noisy cost function \cite{cartis2018improving}. Here, we adopt one of the simplest modifications to increase the number of interpolation points from $(2n + 1)$ to $(n+1)(n+2)/2$ where $n$ is the number of parameters. The modified version, BOBYQA-noisy, gives a much improved result as expected.
Note that the negative measured deviation suggests that the measured energy associated with the optimized parameters is lower than the exact ground-state energy, which is only possible due to the shot noise. This means that the optimization has converged within the limits set by shot noise. (The standard deviation of 100 energy evaluations with 8000 shots is $\sim 0.02$ for all optimized parameters.)

CMA-ES and dual annealing, together with BOBYQA-noisy, have a negative measured deviation for the optimized parameters. (The measured energy is lower than the ground-state energy due to the shot noise.) Among the three optimizers, CMA-ES has the best performance ranked by the error determined by the statevector simulator.
\Cref{fig:optimization_shotnoise_track} suggests that this superior performance is related to the ability to converge to a better solution by CMA-ES than that by BOBYQA-noisy with the cost of a larger number of cost function evaluations. While CMA-ES may still converge to a poor solution for some initial values, a significant portion of the optimization trajectories converges to solutions close to the ground state.
This also suggests the mixed optimization approach is still useful in this noisy case with the BOBYQA-noisy serving as a cross-check of the CMA-ES result.

SPSA shows the best performance among the optimizers being tested. In particular, the error (0.00612) associated with its optimized parameters is even smaller than the noiseless optimization performed by SPSA with an error of 0.04500. This result should not be over-interpreted since these are just two specific optimization trajectories. Nonetheless, this result suggests that the performance of SPSA may be less susceptible to shot noise, and it would be useful to include SPSA to the mixed optimization approach together with other optimizers for VQE with shot noise.

\begin{figure}[t!]
	\centering
	\includegraphics[width=\linewidth]{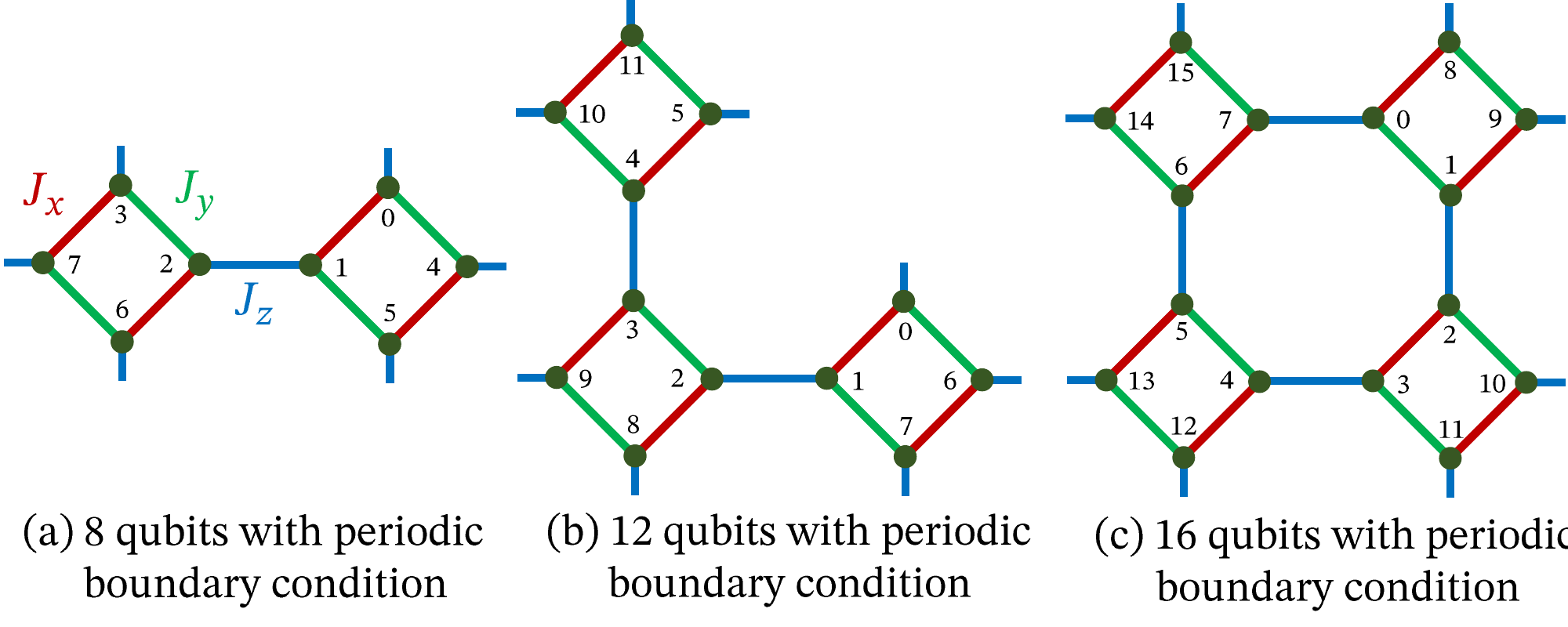}
	\caption{\textbf{
			Lattice geometry with periodic boundary conditions}.  The $J_x$ (red), $J_y$ (green) and $J_z$ (blue) couplings are assigned to each linkage.
		\edit{The qubits at the edge are connected by $J_z$ coupling to the one on the other side of the lattice. In particular, for the periodic boundary conditions, the following qubit pairs are $J_z$ coupled: (a) (0, 5), (3, 6), and (4, 7); (b) (0, 7), (5, 10), (6, 9), and (8, 11); (c) (8, 11), (9, 14), (10, 13), and (12, 15).} 
	}
	\label{fig:periodic_lattice_geometry}
\end{figure}

\begin{table*}
	\begin{center}
		\caption{\textbf{Optimizer performance with shot noise using SPSA}.
			We test the optimizers with various measurement shots and qubits using a one-layer HVA and the (TC$_z$) parameter set. We use the same definition of the error (noiseless) and measured deviation as in \cref{table:optimizer_performance_shotnoise}. The standard deviation of 100 noisy energy evaluations is evaluated at the optimal parameters determined by the optimization. All optimizations have a cutoff of cost function evaluation at 10000.
		}
		\label{table:SPSA_shotnoise}
		\begin{tabular}{ |c|c|c|c|c| } 
			\hline
			Qubits & Shots &  Error (noiseless) &  Measured deviation & Standard deviation of 100 noisy evaluations \\
			\hline
			8 & 8192 & 0.06977 & -0.00246 & 0.00440 \\
			\hline
			12 & 4096 & 0.00191 & 0.00406 & 0.00756 \\
			\hline
			16 & 2048 & 0.14701 & -0.01185 & 0.01103 \\
			\hline
		\end{tabular}
	\end{center}
\end{table*}

\edit{
We further explore the performance of SPSA with various magnitudes of shot noise and lattices shown in \cref{fig:periodic_lattice_geometry} with periodic boundary conditions, and SPSA converges within the shot noise limit.
QPU experiments are more difficult to be carried out with these lattices since the overhead of SWAP gates to simulate the periodic boundary conditions will be substantial for the existing QPU's connectivity. Nonetheless, boundary effects on these lattices are less significant compared to those of the lattices with open boundary conditions. It is thus interesting to investigate numerically how the SPSA performs for these lattices.
We summarize the results in Table~\ref{table:SPSA_shotnoise} associated with one-layer HVA and the TC$_z$ parameter set for the eight-qubit, 12-qubit, and 16-qubit lattices shown in \cref{fig:periodic_lattice_geometry}. We decrease the number of shots for larger lattices since sampling is computationally expensive for large lattices. Similarly, to limit the required computational resources, we fix the cutoff of the number of cost function evaluations to be 10000, which is one tenth of that used in the main text.
In general, the SPSA optimizer converges to a good solution within the limits set by shot noise. The error of the optimized solution determined using a noiseless statevector simulator spans a much wider range of values for different setups.
Further studies with more systematic choices of the number of shots and other lattice and parameter setups will be useful to understand the spread of the errors while SPSA shows a similar level of convergence.
}

\begin{figure*}[t!]
	\centering
	\includegraphics[width=\linewidth]{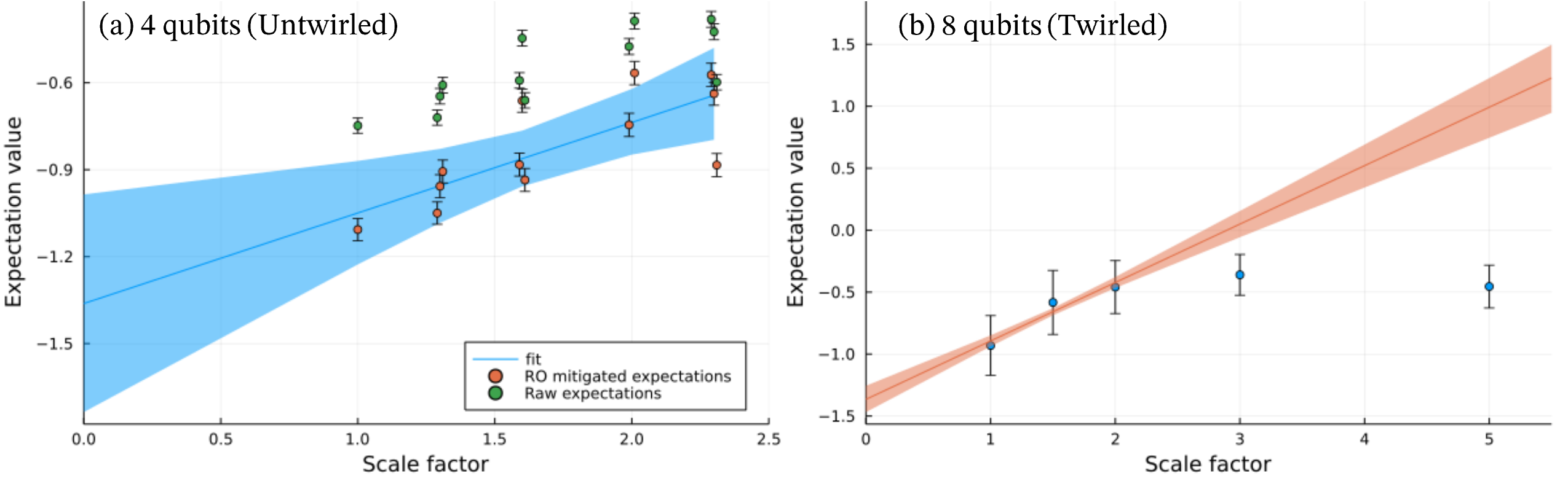}
	\caption{
			\textbf{QPU experiments of optimized HVA circuits (twirled and untwirled) on Aspen-9 with digital ZNE and readout error mitigation}.
			The $x$ axis represents the scale by which the total CZ count in the circuit is increased. QPU data are plotted as the mean and standard deviation of the distribution over expectation values for circuit implementations. A line of best fit is obtained via Bayesian linear regression, and we plot the expectation value of the posterior distribution and a $95\%$ confidence ribbon. The Hamiltonian expectation value, plotted on the $y$ axis, is then extrapolated from the regression to the $x=0$ limit to obtain a fitted noiseless estimate with the depicted uncertainty.
			(a) The four-qubit experiment for various ZNE scale factors uses the HVA-1 ansatz with no (Pauli) twirling. The results extrapolate to roughly $-1.3623$ ($\pm 0.1896$, which is within roughly one standard deviation of the true eigenvalue of $-1.583$). There are one to three circuits for each scale factor (plotted with jitter on the $x$ axis for better visibility) for raw and mitigated expectations, representing several different ways to implement the scaling factor.
			(b) For the eight-qubit experiment, the QPU data points corresponding to scale factors of 3 and 5 are dominated by noise and are not used for ZNE. The results extrapolate to $\sim -1.36$ (with a standard deviation of about $0.054$) which is deviated from the best value of about $-4.23$ that the one-layer HVA can produce.
		}
	\label{aspen_9_expt}
\end{figure*}

\subsection{QPU experiment}

The experimental method has been described in Sec.~\ref{subsecn:expt}.
As discussed there, we perform a proof-of-concept experiment using the HVA with one layer on four-qubit and eight-qubit sub-lattices of Aspen-9 consisting of qubits $(10, 11, 12, 13, 24, 25, 26, 27)$, see Fig. \ref{fig:lattice_geometry}(b). A schematic HVA-1 circuit for eight qubits is depicted in Fig.~\ref{fig:hva-1-circuit}; in order to execute this on Aspen-9, we identify qubits $(q_0, q_1, q_2, q_3, q_4, q_5, q_6, q_7)$ with qubits $(12, 25, 26, 11, 13, 24, 27, 10)$, respectively, on Aspen-9.
To perform digital ZNE, we increase the total CZ count in the circuit by various scale factors and the expectation values measured at each scale factor are mitigated for readout errors.
To improve the results for the eight-qubit case, we twirl the corresponding circuits with random Pauli operators to tailor the noise into a stochastic channel, which may allow ZNE to work better. 

Fig.~\ref{aspen_9_expt} plots the Hamiltonian expectation values in the classically optimized variational state for various scale factors, and the zero-noise limit is extrapolated via Bayesian linear regression. (For the eight-qubit case, the final two scale factors are omitted from the extrapolation because the noise in the system overwhelms any meaningful patterns one could exploit in the extrapolation scheme beyond a scale factor of 2.)
The four-qubit experiment gives the extrapolated energy to be $-1.3623 \pm 0.1896$ which is within one standard deviation of the one-layer-HVA best value $-1.5217$ and, roughly speaking, the true ground-state energy $-1.5831$.
The eight-qubit circuit involves more two-qubit gates and a larger infidelity is expected. Anticipating the larger error to be mitigated, we perform randomized compilation or Pauli twirling to improve the performance of ZNE as explained in Sec.~\ref{subsecn:expt}.
However, even with the additional twirling, we find the extrapolated value for the expectation value to be about $-1.36$ (with a standard deviation of about $0.054$), which is far from the best value of about $-4.23$ that the one-layer HVA can produce in noiseless simulation. This shows that the execution of a $1$-layer HVA circuit for $8$ qubits, which contains $16 \lambda$ CZ gates for scale factor $\lambda \geq 1$, is close to the QPU coherence time. We note that two-qubit gates were executed sequentially in this paper, and parallel execution may lead to an improvement of the results.
The relatively large discrepancy between the optimal value obtained on the QPU versus the exact energy suggests the presence of systematic shifts in the gate parameters when they are executed on the QPU. This may be mitigated when performing the optimization entirely on the QPU. 
Although this is a proof-of-concept experiment, it motivates the use of sophisticated error mitigation techniques along with better gate fidelities to achieve better results on the QPU in the future.

\section{Conclusion}
In this paper, we investigated and benchmarked the appropriate ansatzes and optimization strategies to find the ground state of the square-octagon-lattice Kitaev model with a VQE approach, and demonstrated the ability to determine the system's properties in different parameter regimes using VQE.
We showed that the HVA generally outperforms the two HEAs using CZ or XY gates in the entangling layers. The HEAs require an expensive optimization process due to their rapidly increasing number of parameters with system size and circuit depth, and the unstructured nature of HEAs also makes them susceptible to barren plateaus. The HVA is associated with a relatively easy optimization and is thus better suited to the QPU experiments.
The VQE solutions with HVA allow us to probe static expectation values and the energy gap as a function of Hamiltonian parameters, making it possible to probe the phase diagram using VQE.
We also conducted an experimental test to run classically optimized four-qubit and eight-qubit HVA circuits on the Aspen-9 QPU.
With the help of readout error mitigation and ZNE, the four-qubit experiment gives a result roughly within one standard deviation of the true ground-state energy.
For the eight-qubit experiment, the wider and deeper circuit increases the infidelity of the results. There, we observe a substantial difference from the statevector simulation even with the additional randomized compilation.
This suggests that extra error mitigation techniques and performing the VQE optimization directly on the QPU would be required to extract useful results from the QPU for the future implementation of Kitaev-model VQE experiments.

Multiple optimization algorithms have been tested to optimize the cost function associated with our VQE application. We find that a mixed usage of different optimizers with multiple initial values gives a much more consistent and better result than just using one specific optimizer with one initial value. In particular, a local optimizer (BOBYQA) mixed with a non-local optimizer (CMA-ES) is shown to be an appropriate strategy for VQE calculations of the Kitaev model. SPSA is also a useful addition to the optimization with shot noise.
Although we performed our tests with the noiseless simulator and simulation with shot noise, we expect that QPU experiments will also benefit from this approach as long as the QPU's clock rate is high enough to support the required number of cost function evaluations.

\begin{acknowledgments}
This material is based upon work supported by the U.S. Department of Energy, Office of Science, National Quantum Information Science Research Centers, Superconducting Quantum Materials and Systems Center (SQMS) under contract number DE-AC02-07CH11359.
We would like to thank the entire SQMS algorithms team for fruitful and thought provoking discussions around these works.
We would also like to thank the Rigetti team for assistance with running experiments on the Aspen-9 processor, particularly Matt Reagor, Mark Hodson, Bram Evert, Alex Hill, Eric Hulburd and Dylan Anthony. P.P.O. acknowledges useful discussions with Alexander Huynh and Anirban Mukherjee. Some calculations were performed as part of the XSEDE computational Project No. TG-MCA93S030 on Bridges-2 at the Pittsburgh supercomputer center.
\end{acknowledgments}

\appendix

\bibliography{bibl}

\end{document}